\documentclass[preprint,preprintnumbers,amsmath,amssymb]{revtex4-1}

\usepackage{graphicx}
\usepackage{amsmath}
\usepackage{color}
\usepackage[section]{placeins}
\usepackage{upgreek}
\usepackage{comment}

\newcommand{\be}{\begin{equation}}
\newcommand{\ee}{\end{equation}}
\newcommand{\bfig}{\begin{figure}}
\newcommand{\efig}{\end{figure}}

\begin{document}
\title{Directional dependent Berezinskii Kosterlitz Thouless transition at EuO/KTaO$_3$(111) interfaces}
\author{Zongyao Huang$^{1}$}
\author{Zhengjie Wang$^{1}$, Xiangyu Hua$^{2}$, Huiyu Wang$^1$, Zhaohang Li$^1$, Shihao Liu$^1$, Zhiwei Wang$^1$, Feixiong Quan$^1$,
 Zhen Wang$^1$, Jing Tao$^1$}
\author{James Jun He$^{3}$}
\author{Ziji Xiang$^{2,3}$}
\email{zijixiang@ustc.edu.cn}
\author{Xianhui Chen$^{1,2,3}$}
\email{chenxh@ustc.edu.cn}

\affiliation{
	$^1$ Department of Physics, University of Science and Technology of China, Hefei 230026, China\\
	$^2$ Hefei National Research Center for Physical Sciences at the Microscale, University of Science and Technology of China, Hefei 230026, China\\
	$^3$ Hefei National Laboratory, University of Science and Technology of China, Hefei 230088, China.
}

\date{\today}

\maketitle                   
\noindent
\par
\textbf{In two dimensions, a phase-coherent superconducting state is established via a Berezinskii-Kosterlitz-Thouless (BKT) transition, whose critical temperature $T_{\rm BKT}$ is determined by the global superfluid stiffness in uniform superconducting systems. We report that at the interface between (111)-oriented KTaO$_3$ and ferromagnetic EuO, the two-dimensional superconducting state exhibits a BKT transition relying on the direction of in-plane bias current. The highest $T_{\rm BKT}$ occurs when current is applied along one of the [11$\bar{2}$] axes of KTaO$_3$, underscoring a spontaneous breaking of the threefold lattice rotational symmetry. Such directional dependence of $T_{\rm BKT}$ is consistently reflected in the nonreciprocal signals stemming from superconducting fluctuations above the transition. We attribute this phenomenon to an interfacial phase segregation; the phase with higher $T_{\rm BKT}$ self-organizes into quasi-one-dimensional textures that stretch along one of the [11$\bar{2}$] directions. Our results point toward the emergence of exotic phases of matter beyond the description of conventional BKT physics at a superconducting interface that is subjected to ferromagnetic proximity.
}

\noindent
\textbf{Introduction}
\bigskip

\noindent
In two-dimensional (2D) superconductors ({\it e.g.} thin films and heterointerfaces), the Coulomb interaction energy between a pair of thermally-excited vortex and antivortex depends logarithmically on their separation $r$ until $r$ exceeds the Pearl screening length \textrm{$\Lambda$} = $\lambda^2/d$ (here $\lambda$ is the in-plane London penetration depth, $d$ is the thickness of the superconducting layer) \cite{Pearl}. Thereby, the superconducting transition can be described by the Berezinskii-Kosterlitz-Thouless (BKT) physics \cite{Berezinskii,KT} when \textrm{$\Lambda$} surpasses the size of the 2D superconductor \cite{BMO,Doniach,Minnhagen}; correspondingly, the transition temperature $T_{\rm BKT}$ signifies the dissociation of bound vortex-antivortex pairs at low temperatures ($T$) into liberated vortices, which concomitantly breaks up the superconducting phase coherence. Such a BKT transition does not reflect discontinuities in thermodynamic quantities but signals a universal jump of the 2D superfluid density $n_s$ \cite{Kosterlitz-Nelson}. Within the framework of a 2D XY model, the critical temperature is proportional to the generalized superfluid stiffness $J_{\rm s} \equiv \hbar^2 n_s/4m^*$ ($\hbar$ is the reduced Planck constant and $m^*$ the geometrically-averaged carrier effective mass) \cite{KT,Minnhagen,Kosterlitz-Nelson}:
\begin{equation}
T_{\rm BKT} = \frac{\pi}{2k_B}J_s(T_{\rm BKT}),
 \label{BKT}
\end{equation}
here $k_B$ is the Boltzmann constant.

Superconducting oxide heterointerfaces provide a privileged playground for scrutinizing the BKT physics \cite{Reyren,CavigliaGate,Benfattobroad,BertSQUID,ChangjiangLiu1,ChangjiangLiu2,YanwuXiePRL,HuaSOC,KTOsuperfluid}. They are strongly 2D superconductors with the wave functions of Cooper pairs typically confined within a depth $d_{\rm SC} \lesssim 10$\,nm from the interface between two insulators \cite{Reyren,HuaSOC,YanwuXiePRL}. For SrTiO$_3$ (STO)-based interfaces, e.g., the paradigmatic LaAlO$_3$/SrTiO$_3$ (LAO/STO) \cite{Reyren,CavigliaGate}, the BKT transition is frequently masked by a superconducting percolation process \cite{Benfattobroad,Caprara,VendittiIV} associated with intrinsic spatial inhomogeneity of interfacial electronic states \cite{BertSQUID,Scopigno}; nevertheless, superfluid stiffness measurement validates the BKT scenario for a KTaO$_3$ (KTO)-based interface (in contrast to its STO-based counterparts) \cite{KTOsuperfluid}. More intriguingly, recent studies reveal an extraordinary behavior of the interface between ferromagnetic EuO and (110)-oriented KTO: For samples with 2D carrier density $n_{\rm 2D}$ below a threshold ($\sim$ 8$\times$10$^{13}$ cm$^{-2}$), a higher (lower) superconducting transition temperature $T_{\rm c}$ is invariably observed with the current $I$ applied along the [001] ([1$\bar{1}$0]) axis of KTO, that is, $T_{\rm c}$  appears to be anisotropic with regard to the two orthogonal high-symmetry directions on KTO(110) surface \cite{HuaStripe}. Probably related phenomena have been reported in other KTO-based interfaces with magnetic overlayers \cite{ChangjiangLiu1,Ahadi} or emergent interfacial ferromagnetism \cite{CaZrO3Stoner}, though some of these results suffer from extrinsic geometrical effects \cite{ChangjiangLiu2,HuaStripe}.

Such unusual current-direction dependence of $T_{\rm c}$ strongly challenges the conventional wisdom. According to the BKT theory, a single $T_{\rm BKT}$ is rigorously expected for a uniform 2D superconductor because the global phase coherence of Cooper pairs can only be destroyed all at once. Whereas inhomogeneity engenders superconducting patches with different local $T_{\rm BKT}$ \cite{Benfattobroad}, the resulted percolation process \cite{VendittiIV,Caprara} is not expected to exhibit an anisotropy linking to the crystallographic axes. Hence, a question has been raised --- whether the observed directional-dependent superconducting transitions at KTO-based interfaces can still be understood by assuming a single $T_{\rm BKT}$ \cite{ZixiangLi}?

Here, by performing elaborately designed electrical transport experiments, we demonstrate spontaneous rotational symmetry breaking and directional difference in $T_{\rm BKT}$ at EuO/KTO(111) interfaces. Our results pinpoint the inadequacy of the single (global) $T_{\rm BKT}$ picture; instead, they indicate the formation of quasi-one-dimensional (quasi-1D) channels with a higher $T_{\rm BKT}$ across the interface, which emerge in a self-organized manner.

\bigskip

\noindent
\textbf{Anisotropic BKT transition at EuO/KTO(111) interfaces}
\par
\noindent
We prepared the heterointerfaces by growing a 7-nm-thick EuO film on (111)-oriented KTO single crystal substrates exploiting the molecular beam epitaxy (MBE) technique (Methods); the EuO overlayer is polycrystalline (Supplementary Fig.\,1) owing to the mismatch of lattice constants between two oxides (Methods). The high-symmetry directions at the interface are the [11$\bar{2}$] and [1$\bar{1}$0] crystal axes of KTO, as illustrated in Fig.1a. For KTO-based oxide heterostructures, the interfacial mobile carriers originate primarily from the oxygen vacancies in KTO generated during the overlayer growth \cite{YanwuXiePRL,Zhang2DEG}. As such, we were able to control $n_{\rm 2D}$ of the interfacial 2D electron gas (2DEG) through changing the growth conditions (Methods); samples with $n_{\rm 2D}$ varying over a range of (5.24-9.79)$\times$10$^{13}$ cm$^{-2}$ were attained, all showing superconductivity at low temperatures ($T$) (Supplementary Table\,1). In all these samples, the EuO overlayers exhibit high surface flatness (Supplementary Fig.\,1). An abrupt interface between EuO and KTO is revealed by scanning transmission electron microscopy (STEM) measurements (Supplementary Fig.\,2); the interfacial lattice distortion and the Eu substitution on K sites are confined to the topmost three unit cells of KTO, consistent with earlier reports \cite{ChangjiangLiu1,HuaSOC}, whereas the crystalline state of KTO with well-defined crystallographic directions is preserved up to the interface (Supplementary Note 1 and Supplementary Fig.\,2). The magnetic properties [i.e., a Curie temperature $T_{\rm Curie} \simeq$ 73\,K and an easy-plane magnetic anisotropy (Supplementary Fig.\,3)] are in accordance with previous results \cite{HuaStripe,Zhang2DEG}.

It has been elucidated that the EuO/KTO(110) interfaces with $n_{\rm 2D} \lesssim 8 \times 10^{13}$ cm$^{-2}$ are universally characterized by a concurrence of anisotropic superconducting transitions [$T_{\rm c} (I \parallel [001]) > T_{\rm c} (I \parallel [1\bar{1}0])$] and transport signatures of ferromagnetism (i.e., Anomalous Hall effect and hysteretic magnetoresistance) in the normal-state 2DEGs \cite{HuaStripe}. Occurrence of these captivating phenomena at the EuO/KTO(111) interfaces remains elusive \cite{ChangjiangLiu1,ChangjiangLiu2,MallikARPES}. We first measured the $T$-dependent resistance $R(T)$ on ten EuO/KTO(111) 2DEGs with different $n_{\rm 2D}$ using the van der Pauw (VDP) configuration (Methods and Supplementary Fig.\,4). As shown in Fig.\,1b, eight samples exhibit an anisotropic zero-resistance temperature $T_{\rm c0}$ that is consistently higher when $I$ is along [11$\bar{2}$]; the exceptions are the 2DEGs with higher $n_{\rm 2D}$ ($\geq$ 8.8$\times$10$^{13}$ cm$^{-2}$), which are also characterized by the absence of anomalous Hall effect (Supplementary Fig.\,3c). Therefore, a ubiquitous phenomenology is established for all EuO/KTO interfaces, underscoring the intimate link between directional superconductivity and interfacial ferromagnetism. To preclude the geometrical effects promoted by the VDP configuration \cite{ChangjiangLiu2,HuaStripe}, we patterned \cite{HuaStripe} the EuO/KTO(111) samples into ``L"-shaped Hall-bar devices (Methods, inset of Fig.\,1c) that comprise two channels along [11$\bar{2}$] and [1$\bar{1}$0], respectively. Figures 1c-e displays the resistance data measured on Device 1 ($n_{\rm 2D}$ = 7.67$\times$10$^{13}$ cm$^{-2}$). The normal-state sheet resistance $R_{\rm s}$ exhibit a moderate anisotropy with $R_{\rm s}^{[11\bar{2}]}$ exceeding $R_{\rm s}^{[1\bar{1}0]}$ by 10-15$\%$ over the temperature range of 2-300\,K (Fig.1c). A directional-dependent disparity of superconducting transition is unambiguously observed: the [11$\bar{2}$] channel enters superconducting state at a higher $T$ than the [1$\bar{1}$0] channel, as shown in Figs.\,1d and e.

To further nail down the anisotropic superconductivity, we consider three characteristic temperatures: $T_{\rm c0}$, $T_{\rm BKT}$ and the mean-field critical temperature $T_{\rm c}^{\rm MF}$ (Supplementary Note 4 and Supplementary Fig.\,6). For Device 1, $T_{\rm c0} \approx$ 1.05 (1.15)\,K with $I \parallel$ [1$\bar{1}$0] ([11$\bar{2}$]). $T_{\rm BKT}$ was derived from BKT analysis (Supplementary Note 4) employing the Halperin-Nelson (HN) formula \cite{HNformula} which describes the evolution of paraconductivity (i.e., the excess electrical conductivity prompted by superconducting fluctuations) above $T_{\rm BKT}$; the HN fits (solid lines in Figs. 1d and e) closely track the experimental data, essentially validating the BKT formalism (note that the prerequisite for BKT physics, $L <$ \textrm{$\Lambda$} at $T \rightarrow T_{\rm BKT}$, is satisfied in our devices, see Methods). The HN fits yield $T_{\rm BKT}$ = 1.15\,K and 1.23\,K for $I \parallel $[1$\bar{1}$0] and $I \parallel $[11$\bar{2}$], respectively. A tailed feature deviating from the HN fits occurs at $T \lesssim T_{\rm BKT}$ (Supplementary Note 4 and Supplementary Fig.\,6) and is attributed to the impact of spatial inhomogeneity \cite{Benfattobroad,BenfattoLa214}. Estimation of $T_{\rm c}^{\rm MF}$ relies on two approaches: (i) fits of the paraconductivity to the Aslamazov-Larkin model (Supplementary Fig.\,6) and (ii) position of the inflection point in $R(T)$ (insets of Figs.\,1d and e) \cite{ChangjiangLiu2,BenfattoLa214}; the coincidence between (i) and (ii) gives $T_{\rm c}^{\rm MF}$ = 1.17(1.25)\,K when $I \parallel$ [1$\bar{1}$0] ([11$\bar{2}$]). The directional dependence of all three characteristic temperatures (with a consistent disparity $\Delta T \simeq$ 0.1\,K between [1$\bar{1}$0] and [11$\bar{2}$]) provide definitive evidence for the intrinsically anisotropic superconductivity at the EuO/KTO(111) interface. In particular, the anisotropy of $T_{\rm BKT}$ marks a clear departure from the conventional BKT picture that underlines the characteristic energy scale of global phase stiffness. We also note that the upper critical field $H_{\rm c2}$ measured on Device 1 depends correspondingly on the direction of $I$ (Supplementary Fig.\,5), resembling the behavior of the EuO/KTO(110) 2DEGs \cite{HuaStripe}.

\bigskip

\noindent
\textbf{Spontaneous breaking of crystalline rotational symmetry}
\par
\noindent
In contrast to the KTO(110) interfaces which are inherently twofold symmetric, the (111) surface of cubic KTO exhibits threefold lattice symmetry: there are three equivalent crystal axes for both [11$\bar{2}$] and [1$\bar{1}$0] directions (Fig.1a); meanwhile, the 2D electronic structure adopts a sixfold ($C_{\rm 6}$) rotational symmetry \cite{BrunoARPES}. Such difference offers a good opportunity for examining the correspondence between directional BKT transition and the lattice symmetry. A superconducting stripe model has been proposed for the EuO/KTO(110) interfaces; it postulates the emergence of quasi-1D textures harboring more resilient superconductivity and running along the [001] direction \cite{HuaStripe,ZixiangLi}. Applying this scenario to the (111) interfaces, a spontaneous breaking of the $C_{\rm 6}$ rotational symmetry in transport properties (inherited from the 2DEG Fermi surface \cite{BrunoARPES}) is expected: stripes formed along three [11$\bar{2}$] directions must cross, resulting in connected paths in all in-plane directions and hence eliminating directional signatures. Alternatively, if the directional-dependent superconducting phenomena stem from Fermi surface properties \cite{Ahadi}, the $C_{\rm 6}$ rotational symmetry should be maintained.

We investigated this issue by fabricating a ``double-tri-beam" device on a EuO/KTO(111) 2DEG (Device 2, $n_{\rm 2D}$ = 5.24$\times$10$^{13}$\,cm$^{-2}$) (Methods). It contains two parts, each consists of three Hall-bar channels arranged in a tri-beam configuration along three [11$\bar{2}$] or [1$\bar{1}$0] directions, respectively (insets of Figs.\,2a and b). As displayed in Fig.2a, $R(T)$ measured on three [11$\bar{2}$] channels verify that the $C_{\rm 6}$ rotational symmetry is broken: superconductivity develops in one channel (marked by cyan color in the inset) at a substantially higher $T_{\rm BKT}$ compared to the other two channels. Note that this [11$\bar{2}$] direction is not the one parallel to the edge of substrate (navy). Concurrently, the lowest $T_{\rm BKT}$ occurs in the channel (colored in red) along a [1$\bar{1}$0] direction that is perpendicular to the [11$\bar{2}$] channel showing the highest $T_{\rm BKT}$. Consistent observations are also obtained from two ``radial" six-beam devices, Device 4 and Device 5 (Supplementary Fig.\,7); these devices are composed of six Hall-bar channels (three [11$\bar{2}$] and three [1$\bar{1}$0]) radiating from the same spot on the sample, a design aiming at minimizing the influence of spatial inhomogeneity. In both devices, there is one [11$\bar{2}$] channel exhibiting the highest $T_{\rm BKT}$, whilst the [1$\bar{1}$0] channel orthogonal to it displays the lowest $T_{\rm BKT}$ (Supplementary Fig.\,7 and Table\,4). These results congruently point towards a preponderant twofold rotational symmetry for the anisotropic superconductivity, which explicitly contradicts the crystalline anisotropy. Twofold anisotropic superconducting states at KTO(111) interfaces has been revealed under magnetic fields ($H$) and assigned to mixed-parity superconducting pairing \cite{WeiLiRotation}; here, the anisotropy is manifested in the BKT transition at zero $H$, defying straightforward understanding based on whatever pairing symmetry.

An analysis of the $I-V$ characteristics (Supplementary Note 5) bring more insights into such directional BKT transitions. Within the BKT theory \cite{Minnhagen,HNformula}, a nonlinear $I-V$ relationship emerges at $T_{\rm BKT}$: $V \propto I^{\alpha}$, $\alpha = 1+\pi J_{\rm s}/k_B T$; $\alpha$ = 3 at $T = T_{\rm BKT}$ and 1 for all $T > T_{\rm BKT}$. In Fig.\,2c we plot $\alpha(T)$ measured on the pair of orthogonal [11$\bar{2}$] and [1$\bar{1}$0] channels with maximum and minimum $T_{\rm BKT}$, respectively (see Supplementary Fig.\,8 for raw data). Using the $\alpha$ = 3 criterion, $T_{\rm BKT}$ is determined to be 1.40\,K for the [11$\bar{2}$] channel and 1.16\,K for the [1$\bar{1}$0] channel, in agreement with HN fitting results (Figs.\,2a and b). Effect of inhomogeneity smears the sharp jump of $\alpha$ (and hence $n_{\rm s}$, Eq.\,\ref{BKT}) at $T_{\rm BKT}$ \cite{Benfattobroad,BenfattoLa214}, yet its influence is weaker compared with the LAO/STO interfaces \cite{VendittiIV}; the BKT description thus stays valid in the present instance (Supplementary Note 4 and 5). The anisotropy in $\alpha$ is significant: at $T_{\rm BKT}$($I \parallel [11\bar{2}]$) = 1.40\,K, $\alpha$ is approximately 1 along [1$\bar{1}$0] (Fig.\,2c). Between these two $T_{\rm BKT}$s, current-induced unbinding of vortex-antivortex pairs can only occur along [11$\bar{2}$] but not along [1$\bar{1}$0] (thermally-activated dissociation appears to be realized in this direction). The intrinsic essence of anisotropy is further verified by its dependence on current amplitudes. As illustrated in Fig.\,2d, a persistent disparity of $T_{\rm c0}$ survives at the zero-bias limit, excluding current-induced nonequilibrium effects and spurious measurement issues as the origin of the directional behaviors.

\bigskip

\noindent
\textbf{Second-order nonreciprocal charge transport}
\par

\noindent
As the oxide heterointerfaces implicitly breaks inversion symmetry, the interfacial 2DEGs possess spin-split Fermi surfaces provoked by Rashba-type spin-orbit coupling (SOC) \cite{BrunoARPES,HeBLMR}. A notable outcome of such Rashba spin splitting is the nonreciprocal charge transport under $H$-fields: $\Delta R = R(I)-R(-I) \neq 0$. For an interfacial 2DEG, the nonreciprocity is given by $\Delta R = \mu_0\gamma \vec{I}\cdot(\hat{P}\times\vec{H}$), where $\gamma$ is a coefficient characterizing the current rectification and $\hat{P}$ is a unit vector normal to the interface \cite{ChoeLAO-STO,ItahashiSTO,ZhangLightInduced}. Nonreciprocal transport can be detected in both DC \cite{ChoeLAO-STO,ZhangLightInduced} and second harmonic \cite{HeBLMR,ItahashiSTO} measurements; it serves as a powerful tool for exploring the superconducting fluctuations heralding the BKT transition \cite{ItahashiSTO,WakatsukiMoS2,Hoshino}. Here we opted for a directional measurement of second-harmonic resistance $R^{2\omega}$ (configuration shown in Fig.\,3a, inset) employing an AC lock-in technique (Methods). An in-plane $\mu_0H$ = 0.2\,T was applied perpendicular to the biased channel in a Hall-bar device (Device 3, $n_{\rm 2D} = 5.59\times10^{13}$ cm$^{-2}$). The amplitude of applied AC current (1 $\mu$A) is within the linear-response regime ($R^{2\omega}$ $\propto$ $I$), as shown in Supplementary Fig.\,10d. $\gamma = \frac{2R^{2\omega}}{\mu_0 HI R^{\omega}}$ was derived from the ratio between $R^{2\omega}$ and the linear resistance $R^{\omega}$ \cite{HeBLMR,ItahashiSTO,WakatsukiMoS2}. Fig.\,3a shows a significant enhancement of $\gamma$ in the superconducting fluctuation regime \cite{WakatsukiMoS2}, giving rise to a salient peak feature (with gigantic amplitudes exceeding 4$\times$10$^4$\,T$^{-1}$A$^{-1}$) slightly above $T_{\rm BKT}$. A $(T-T_{\rm BKT})^{-3/2}$ divergence of $\gamma$ is observed as $T \rightarrow T_{\rm BKT}+$ (Figs.\,3b and c), in agreement with theoretical predictions \cite{ItahashiSTO,Hoshino}. Further analysis allow us to separate the contributions from amplitude fluctuations above $T_{\rm c}^{\rm MF}$ and vortex motion between $T_{\rm c}^{\rm MF}$ and $T_{\rm BKT}$ (Supplementary Note 6). It can be seen that $\gamma(T)$ evolves qualitatively conformably for $I$ along [11$\bar{2}$] and [1$\bar{1}$0], whereas the deduced characteristic temperatures $T_{\rm BKT}$ (1.00 versus 1.10\,K) and $T_{\rm c}^{\rm MF}$ (1.12 versus 1.20\,K) are consistently higher (lower) with $I \parallel [11\bar{2}]$ ([1$\bar{1}$0]); moreover, $\gamma$ is larger along [1$\bar{1}$0] in the vortex-flow regime albeit the anisotropy becomes weak above the higher $T_{\rm c}^{\rm MF}$ (Figs.\,3b and c, Supplementary Fig.\,9).

More perplexing signatures were observed in dependence of $R^{2\omega}$ on an in-plane $H$. Data presented in Fig.\,3d and e reveal that the temperature range showing enhanced $R^{2\omega}$ can be divided into three regions, determined from $R^{2\omega}(H)$ (here we refer to its behavior under positive $H$): (i) Above $T_{\rm BKT}(H = 0)$, $R^{2\omega}$ (and thus $\gamma$) is negative at the low-$H$ limit, developing a downward skewed-peak feature before evolving into a weak positive value at a field $H_{\rm N}$; (ii) when $T \simeq T_{\rm BKT}(H = 0)$, a positive component of $R^{2\omega}$ emerges in the intermediate range of $H$ and overcomes the negative signal between  two characteristic fields $H^*_1$ and $H^*_2$; (iii) for $T \leq$ 0.8\,K, the low-$H$ negative signal disappears and $R^{2\omega}$ is essentially zero up to $H = H^*_0$, meanwhile the positive signal between $H^*_0$ and $H^*_2$ evolves into a sharp peak (more pronounced along [11$\bar{2}$]) with decreasing $T$. The $H-T$ phase diagrams depicted as contour plots in Fig.\,4 further illustrate such sophisticated sign reversal. $H_{\rm N}$, $H^*_0$ and $H^*_2$ increase upon cooling as generally expected for characteristic fields of a superconductor; conversely, $H^*_1$ shows the opposite trend by reaching zero at $\simeq$ 0.9(1.0)\,K with $I \parallel [1\bar{1}0]$ ([11$\bar{2}$]) (Figs.\,4a and b). Thereby, the positive $R^{2\omega}$ signal (marked by red at $H >$ 0) occurs in a wing-shaped regime outlined by the ``M"-like upper boundary composed of $H^*_1(T)$ and $H^*_2(T)$ together with the concave lower boundary delineated by $H^*_0(T)$. More intriguingly, there are notable coincidences between such characteristic fields for $R^{2\omega}$ (Figs.\,4a and b) and contour lines of linear $R = R^{\omega}$ (Figs.\,4c and d): $H_{\rm N}(T)$ (below $T \sim$ 1.5\,K), $H^*_2(T)$ and $H^*_0(T)$ roughly trace the lines of $R$ = 0.8, 1/3 and 0.01\,$R_{\rm N}$ (the normal-state resistance), respectively. These correspondences corroborate that the sign reversal of $R^{2\omega}$ cannot be associated with the vortices induced by a small out-of-plane $H$ (due to sample misalignment) \cite{Masuko}; we alternatively established a model based on the renormalization of the superfluid density \cite{Hoshino} to interpret the sign-changing profile of nonreciprocity (Supplementary Note VI). (Similar phenomena under in-plane $H$ are observed in other spin-momentum-locking 2D superconductors \cite{Bi2Te3-FeTe}.) In closing, we remark the central implication of Fig.\,4 --- almost all the characteristic fields and temperatures are 1-10$\%$ higher for the [11$\bar{2}$] channel compared to [1$\bar{1}$0] (with the only exception of $H^*_0$ at 0.6\,K), underscoring universally directional-dependent pertinent energy scales of the interface superconductivity.

\bigskip

\noindent
\textbf{\large Discussion}
\par

\noindent
Our results confirm the directional dependence of BKT transition at the EuO/KTO(111) interfaces. The highest $T_{\rm BKT}$ clings to one [11$\bar{2}$] direction, inherently breaking the lattice rotational symmetry. It would be seen at odds with the conventional formalism $T_{\rm BKT} \propto J_s(T_{\rm BKT})$ (Eq.\,\ref{BKT}), wherein $J_{\rm s}$ is a global character and hence $T_{\rm BKT}$ is a scalar. One way to reconcile the ostensible anisotropy of $T_{\rm BKT}$ with a uniform 2D conductivity is to postulate a directional-dependent vortex core energy $\mu$. In principle, if $\mu \neq \mu_{\rm XY}$ ($\mu_{\rm XY} \simeq \pi^2J_s/2$ \cite{KT,Minnhagen}), $T_{\rm BKT}$ deviates from the universal position predicted by 2D XY model, i.e., Eq.\,\ref{BKT} \cite{BenfattoLa214,BenfattoCoreEnegy,YongNbN}; a larger $\mu$ with $I \parallel$ [11$\bar{2}$] renders $T_{\rm BKT}$ higher along this direction. However, our HN analysis (Supplementary Note 4) defies such an interpretation (Supplementary Table\,3). We thus suggest that the superconducting interfaces studied herein can be neither uniform (as also hinted by the small portion of $n_s/n_{\rm 2D} \sim 2-10 \%$ \cite{BertSQUID,KTOsuperfluid}) nor randomly inhomogeneous without specific orientational preference \cite{Benfattobroad,Caprara,VendittiIV}; in particular, the $I-V$ characteristics lack evidence for an arbitrary array of Josephson junctions (Supplementary Note 5).

The superconducting stripe model \cite{HuaStripe} explains the data more naturally: The interfacial 2DEG self-organized into two spatially separated phases, the phase with higher $T_{\rm BKT}$ forms quasi-1D channels stretching along [001] and the specific [11$\bar{2}$] direction at the EuO/KTO(110) \cite{HuaSOC} and (111) interfaces, respectively. Several further comments could be made about this picture. First, the stripes are likely to be meandering ``rivers" \cite{ZixiangLi} rather than straight lines, leading to contingent intermediate $T_{\rm BKT}$ when $I$ is not along or perpendicular to such specific directions (Figs.\,2a and b); this is corroborated by the results measured on six-beam Device 4 (Supplementary Table 4). Second, the superconducting transition along the high-$T_{\rm BKT}$ [11$\bar{2}$] direction cannot be described by the Langer-Ambegaokar-McCumber-Halperin theory (Supplementary Fig.\,11), invalidating a superconducting nanowire description \cite{CaZrO3Stoner} of the stripes (Supplementary Note 7). This means the widths of stripes must exceed the in-plane coherence length ($\xi \sim$ 17-22\,nm, Supplementary Note 3). Third, the Josephson coupling between neighboring stripes ought to be substantially suppressed to ensure the occurrence of two individual $T_{\rm BKT}$s. A recent proposal considers the higher-order inter-stripe coupling which prompts different types of vortex excitations (exhibiting distinct $I-V$ characteristics) along the two directions, subsequently endows the system with directional-dependent apparent $T_{\rm c0}$ \cite{ZixiangLi}. However, such model predicts a single $T_{\rm BKT}$ and a vanishing disparity of $T_{\rm c0}$ at the $I$ = 0 limit, whereas our data contradict both (Figs.\,2c and d). The compromise between phase nonuniformity and Josephson coupling await further investigations.

It is worth noting that according to the superconducting stripe interpretation, the stripe phase has different onset temperatures and characteristic directions between EuO/KTO(110) and (111) interfaces. This is likely to be associated with the distinct 2DEG electronic structures for these two orientations of KTO \cite{HuaStripe,MallikARPES,LAO-KTOARPES}. For KTO-based superconducting interfaces, orientation-dependent optimal $T_{\rm c}$ has been established \cite{YanwuXiePRL,ChangjiangLiu2} with the hierarchy $T_{\rm c}(100) < T_{\rm c}(110) < T_{\rm c}$(111), which originates from the orientation dependence of (Ta 5$d$ $t_{\rm 2g}$) orbital degeneracy and electron-phonon coupling strength for the interfacial 2DEG \cite{ChangjiangLiu2,LAO-KTOARPES}. The same combination probably also rules the onset temperature of superconducting stripes (which relies on the pairing instability as well), rendering it higher at the (111)-oriented interfaces. The characteristic directions of stripes are putatively linked to the detailed electronic band structures of 2DEGs, yet a comprehensive picture is lacking thus far. We stress that the superconducting stripe scenario is still hypothetical at this stage and is awaiting further verification. 

The concurrence of directional-dependent superconducting transition and interfacial ferromagnetism --- unmasked for both (110)- and (111)-oriented EuO/KTO interfaces --- not only precludes pure structural origins (e.g., ferroelastic domain walls \cite{DomainWall}) of the putative superconducting stripes, but also signifies the influence of unconventional superconducting pairing between spin-polarized electrons. Whilst the ferromagnetic KTO-based 2DEGs (with an easy-plane anisotropy) can emerge from either proximity effect of magnetic overlayers (EuO for instance) \cite{HuaSOC,Zhang2DEG} or Stoner-type instability of Ta 5$d$ electrons in nonmagnetic heterostructures \cite{CaZrO3Stoner,WeiLiFM}, we speculate that anisotropic $T_{\rm c}$ is an ubiquitous signature of these 2DEGs regardless of the mechanism for ferromagnetic order \cite{HuaStripe,Ahadi,CaZrO3Stoner,MallikARPES}. In particular, the [1$\bar{1}$0] direction universally characterizes a lower $T_{\rm c}$ for both KTO(111) and (110) interfaces in all these cases. A combination of strong Rashba SOC and spontaneous in-plane Zeeman field potentially promotes finite-momentum pairing \cite{HuaSOC,PALee} or $p$-wave pairing \cite{WeiLiRotation,WeiLiFM,Kozii} at the interface; these unconventional superconducting states, however, do not promise the observed directional dependence by their own. We have also noticed that evidence for directional-dependent superconducting transitions has recently been unveiled in nonmagnetic LAO/KTO heterostructures \cite{WuNematicity} and infinite-layer nickelate thin films \cite{LiNSNO} (yet both develop an arbitrary symmetry axis that is unrelated to any high-symmetry crystallographic directions); it is presently unknown whether there are underlying connections among all these cases. Further elucidation of such emergent physics can be a crucial step towards a unified picture for 2D systems with coexisting ferromagnetism and superconductivity.

\bigskip

\noindent
\textbf{\large Methods}

\noindent
\textbf{Film growth}

\noindent
EuO thin films were epitaxially grown on (111)-oriented KTO single crystals using a molecular beam epitaxy (MBE) system with a base pressure of 4$\times$10$^{-10}$ mbar. The typical in-plane dimensions of the KTO substrates are 5$\times$5\,mm$^{2}$. Prior to the growth process, the substrates were pre-annealed at 600\,$^\circ$C for one hour and subsequently cooled to the deposition temperature. The deposition rate (calibrated by a quartz-crystal monitor) of Eu was set to 0.3 \AA/s; supply of oxygen was turned on after 10 minutes of Eu deposition. After growth, the samples were cooled to room temperature with the oxygen supply turned off. A protective germanium layer (3–4 nm in thickness) was deposited at room temperature to prevent oxidation. For KTO-based heterostructures, it has been established that the interfacial two-dimensional electron gas (2DEG) arises predominantly from the oxygen vacancies \cite{YanwuXiePRL,YanwuXiePRL2}; in particular, since the deposited Eu extracts oxygen from KTO to form the EuO overlayer, the oxygen vacancies created in (the first few layers of) KTO promote the 2DEG at the EuO/KTO interfaces \cite{Zhang2DEG}. The primary strategy for controlling the carrier density $n_{\rm 2D}$ of the interfacial 2DEG is to tune the oxygen partial pressure and temperature during growth. In contrast to a previous work on KTO(110) interfaces \cite{HuaStripe}, here we figured out that a reduction of the oxygen partial pressure leads to higher $n_{\rm 2D}$; meanwhile, higher growth temperatures are required for achieving higher $n_{\rm 2D}$. The growth conditions of the samples studied in this work are listed in Supplementary Table\,1. The typical thickness of EuO overlayer is 6-7 nm.

We note that the EuO overlayers in the EuO/KTO(111) heterostructures are polycrystalline owing to the large lattice mismatch between the two oxides. Both EuO and KTO crystallize in a cubic structure, with the lattice constants $a_{\rm EuO}$ = 5.150\,\AA, and $a_{\rm KTO}$ = 3.989\,\AA, respectively. As shown in Supplementary Fig.\,1a, the nearest Ta-Ta atomic distance are 5.640\,\AA\, along the [1$\bar{1}$0] direction and 9.770\,\AA\, along the [11$\bar{2}$] direction on the KTO (111) surface; as such, an anisotropic strain larger than 5\,$\%$ is sensed by the EuO overlayer, engendering its polycrystalline epitaxy as confirmed by the ring-shaped reflective high-energy electrons diffraction (RHEED) pattern (Supplementary Fig.\,1c) and XRD measurement (Supplementary Fig.\,1b). The surface flatness of the samples were examined by atomic force microscopy (AFM) measurements. It was determined that the root mean-squared roughness of the EuO films (Supplementary Figs.\,1d and e) and KTO substrates (Supplementary Fig.\,1f) is 0.35-0.44 nm and $\simeq$ 0.27\,nm, respectively (these values translate to less than two atomic layers of (111)-oriented KTO); no unidirectional step-and-terrace surface structure can be observed on the annealed KTO substrates (Supplementary Fig.\,1f).

\noindent
\textbf{Device fabrication}

\noindent
Fabrication of the Hall-bar devices exploited a combination of conventional photolithography and the argon-ion-beam etching technology. The heterointerface samples first went through an ultrasonic cleaning process in acetone, followed by spin-coating of photoresist (5000 rpm for 60 seconds); to achieve proper resist adhesion, a soft-bake procedure was performed using a hot plate, with samples heated to 100\,$^\circ$C for 90 seconds. The samples were then exposed to ultraviolet for 10 seconds with specific shadow masks attached. Subsequently, they were immersed in diluted developer solution and deionized water for 1 minute to remove the photoresist outside the designed areas of devices, before being transferred into the argon-ion etching chamber. After an etching process of 15 minutes, an effective etch depth of approximately 60\,nm can be achieved, which far exceeds the thickness of EuO films (6-7 nm). The residual photoresist on the device areas was eventually removed by acetone. 

In this study, we fabricated devices with four different designs on six samples: (i) Standard Hall-bar patterns on Sample No.\,1 (Device 1, main text Fig.\,1c); (ii) double ``tri-beam" patterns on Sample No.\,2 (Device 2, main text Fig.\,2), (iii) the ``radial" six-beam pattern on Samples No.\,8 and No.\,9 (Devices 4 and 5, respectively, Supplementary Fig.\,7); (iv) miniature Hall-bar patterns on Sample No.\,3 (Device 3 for nonreciprocal transport experiments, Supplementary Fig.\,9a) and Sample No.\,10 (Device 6, Supplementary Fig.\,10d). For the Hall-bar channels in (i), (ii) and (iii), (iv) the effective sizes (length $L \times$ width $W$) are 700$\times$100, 600$\times$100 and 60$\times$10 $\mu$m$^2$, respectively; the reduced size of device type (iv) ensures a lower noise level for the second harmonic resistance measurements. We mention that these sample sizes do not overcome the Pearl screening length of the interfacial superconducting state, thus guarantee the validity of BKT scenario: For a rough estimation of the Pearl screening length, we consider the superfluid stiffness measured for the AlO$_x$/KTO interface \cite{KTOsuperfluid}: $J_{\rm s} \simeq$ 7.3\,K at the zero-$T$ limit, which converts to \textrm{$\Lambda$} = $\frac{\hbar^2}{4\mu_0 e^2 k_B}\frac{1}{J_{\rm s}} \simeq$ 860\,$\mu$m at $T$ = 0\,K ($\mu_0$ is the vacuum permeability); this value is of the order of the lengths of Hall-bar devices ($L$ = 600-700\,$\mu$m, hence $L <$ \textrm{$\Lambda$} is validated at $T \rightarrow T_{\rm BKT}$.

\noindent
\textbf{Electrical transport measurement}

\noindent
The transport measurements were carried out in a Quantum Design Physical Property Measurement System (PPMS-9\,T) equipped with a dilution refrigerator insert. Aside from the Hall-bar devices, measurements of longitudinal and Hall resistance were also performed on all samples using a van der Pauw configuration \cite{HuaStripe}. Electrical contacts were made by the aluminium wire bonding technique. For in-plane magnetic field application ($H$ parallel to the sample surface), the sample was attached to a rectangular-shape copper block with its surface perpendicular to the PPMS puck. Second harmonic resistance was measured employing an ac lock-in technique. An ac current with the frequency $f$ = 17.777 Hz was supplied to the devices; the first and second harmonic signals of the ac voltage were simultaneously collected by lock-in amplifiers (Stanford Research Systems Model SR830 DSP). During the measurements, we kept the phases of first-harmonic and second-harmonic components at $\simeq$ 0 and $\simeq$ -$\pi$/2, respectively.

\bigskip
\noindent
\textbf{Data availability:} Data for all graphs presented in this paper are available from the authors upon request.


\bigskip

\bigskip




\noindent
\textbf{Acknowledgements}

\noindent
We thank Kun Jiang, Zi-Xiang Li, Yi Zhou, Hong Yao and Shunqing Shen for fruitful discussions. This work was supported by the National Natural Science Foundation of China (Grants Nos. 12488201, 12274390, 12304035 and 12204451), the National Key Projects for Research and Development of China (Grant No. 2022YFA1602602), the Quantum Science and Technology-National Science and Technology Major Project (2021ZD0302800), the Chinese Academy of Sciences Superconducting Research Project under Grant No. SCZX-0101 and the Basic Research Program of the Chinese Academy of Sciences Based on Major Scientific Infrastructures (No. JZHKYPT-2021-08).
\bigskip

\noindent
\textbf{Author contributions}

\noindent
Z.H., Z.X. and X.C. conceived the research project. Z.X. designed the experiments.  Z.H. and Z.-J.W. prepared the heterointerface samples. Z.H. and X.H. fabricated the Hall-bar devices with assistance from S.L. Z.H., X.H. and Z.-J.W. performed the transport measurements with the help from Z.L., S.L., Z.-W.W. and F.Q. H.W., Z.W. and J.T. conducted the scanning transmission electron microscopy experiments. J.J.H., Z.X. and X.C. analyzed the data. J.J.H. provided theoretical modeling of the second harmonic resistance. Z.H. and Z.X. wrote the manuscript with inputs from all authors.
\bigskip

\noindent
\textbf{Competing interests}

\noindent
The authors declare no competing interests.
\bigskip

\noindent
\textbf{Correspondence} and requests for materials should be addressed to Ziji Xiang or Xianhui Chen.


%

\begin{figure*}[htbp!]
	\centering
	\includegraphics[width=1\textwidth]{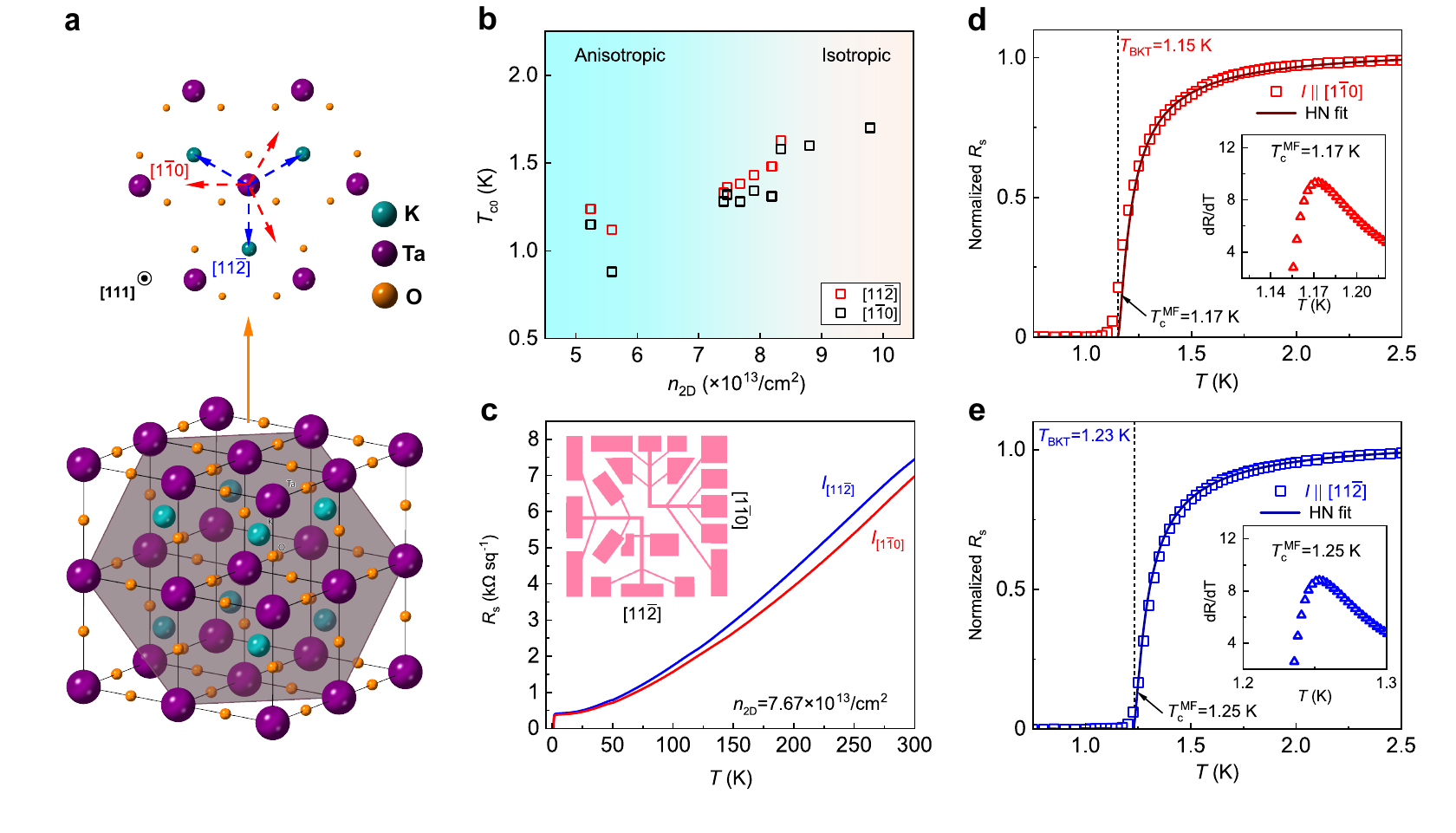}
	\caption{\textbf{Transport measurements on 2D electron gas (2DEG) at the EuO/KTaO$_3$ (111) interfaces.} \textbf{a}, A schematic illustration representing the lattice structure of cubic KTaO$_3$; an expanded view of the (111) plane is shown with the in-plane high-symmetry directions marked by dashed arrows: three equivalent [11$\bar{2}$] axes (blue) and three equivalent [1$\bar{1}$0] axes. \textbf{b}, Superconducting transition temperature $T_{\rm c0}$ measured in a van der Pauw geometry on ten interface samples (see Supplementary Fig.\,4 for raw data), plotted against the carrier density $n_{\rm 2D}$ of the 2DEGs determined from Hall data (details are listed in Supplementary Table 1). $T_{\rm c0}$ is defined to be when the sheet resistance $R_{\rm s}$ reaches 0.1$\%$ of its normal state value $R_{\rm N}$ at $T$ = 3\,K. Red squares and black circles denote $T_{\rm c0}$ measured with current $I$ applied along the crystallographic [11$\bar{2}$] and [1$\bar{1}$0] directions, respectively. Blue and yellow shaded area highlights the presence and absence of the directional disparity of $T_{\rm c0}$, respectively. \textbf{c}, $R_{s}$ as a function of $T$ measured on the Hall-bar Device 1 (illustrated in the inset) which permits a simultaneous detection of $R_{s}$ along the [11$\bar{2}$] (blue) and [1$\bar{1}$0] (red) channels. \textbf{d},\textbf{e}, Low-$T$ $R_{\rm s}$ ($T <$ 2.5\,K) signifying the superconducting transitions for $I \parallel $[1$\bar{1}$0] (\textbf{d}) and [11$\bar{2}$] (\textbf{e}), respectively. Solid lines are fits to the Halperin-Nelson (HN) formula; vertical dotted lines indicate the yielded $T_{\rm BKT}$ (BKT transition temperature). Insets show $dR_{\rm s}/dT$ versus $T$ and its maximum (i.e., the inflection point) implies the position of mean-field critical temperature $T_{\rm c}^{\rm MF}$ (black arrows) \cite{BenfattoLa214}.
	}
	\label{Fig1}
\end{figure*}

\begin{figure*}[htbp!]
	\centering
	\includegraphics[width=0.7\textwidth]{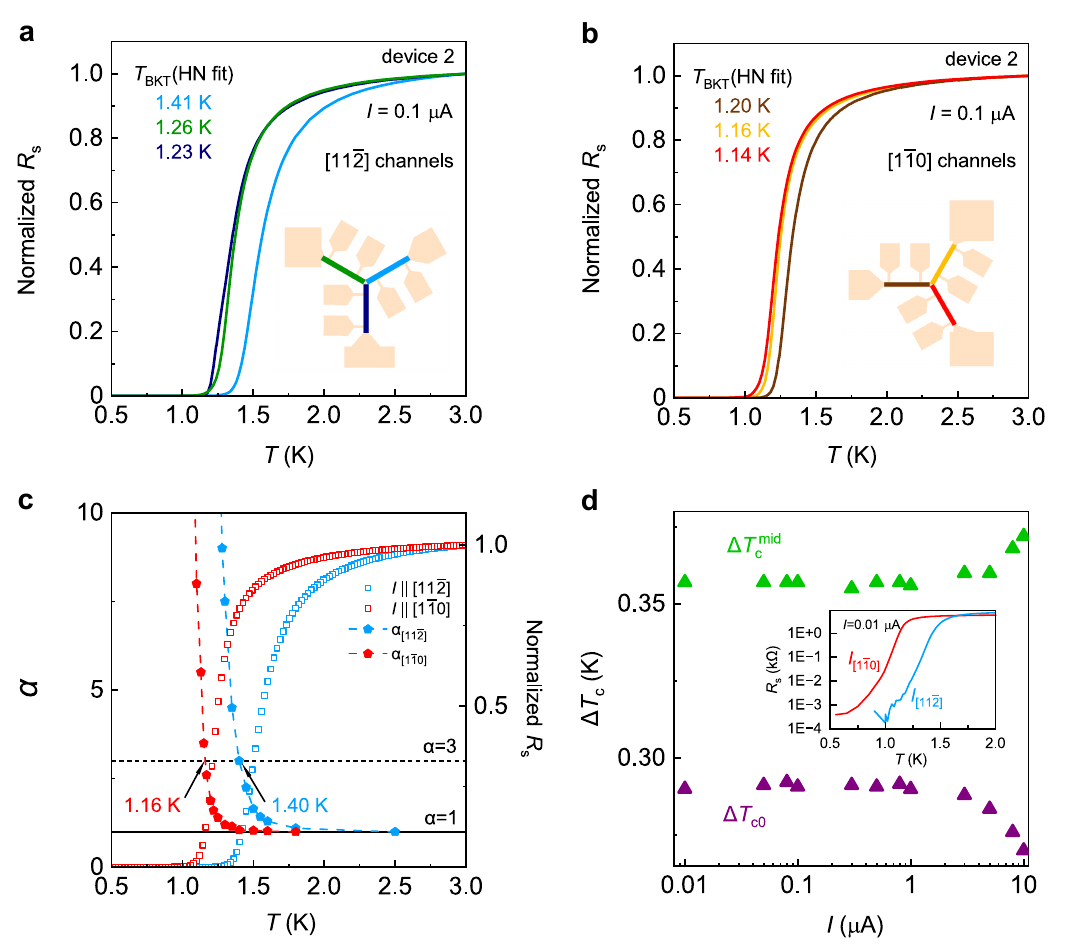}
	\caption{\textbf{Broken in-plane rotational symmetry in the superconducting state.} \textbf{a},\textbf{b}, Normalized temperature-dependent resistance $R(T)$ curves showing different superconducting transition temperatures along three equivalent [11$\bar{2}$] (\textbf{a}) and [1$\bar{1}$0] (\textbf{b}) directions, respectively. Data were measured on Device 2 with the carrier density $n_{\rm 2D}$ = 5.24$\times$10$^{13}$ cm$^{-2}$); the [11$\bar{2}$] and [1$\bar{1}$0] Hall-bar channels (colored correspondingly to the curves) are illustrated in the insets. $T_{\rm BKT}$ (BKT transition temperature) attained from Halperin-Nelson (HN) formula fits (Supplementary Fig.\,6) are noted beside the curves. The maximum (minimum) $T_{\rm BKT}$ = 1.41 (1.14)\,K is observed for the [11$\bar{2}$] ([1$\bar{1}$0]) channel marked by cyan (red) color in \textbf{a} (\textbf{b}); these two channels are orthogonal to each other. \textbf{c}, Temperature dependent exponent $\alpha(T)$ extracted from the $I-V$ (current-voltage) characteristics: $V = I^{\alpha}$ (Supplementary Fig.\,8). Blue and red symbols denote the data for the [11$\bar{2}$] channel with the highest $T_{\rm BKT}$ in \textbf{a} and the [1$\bar{1}$0] channel with the lowest $T_{\rm BKT}$ in \textbf{b}, respectively; the $I-V$ analysis yields a $T_{\rm BKT}$ (at which $\alpha$ = 3) 0.01-0.02\,K lower than the HN fitting result for each direction. \textbf{d}, The directional disparity of superconducting transitions, $\Delta T_{\rm c}$, plotted against the applied current $I$ for the two channels referred to in \textbf{c}. $\Delta T_{\rm c}$ were calculated for $T_{\rm c0}$ and $T_{\rm c}^{\rm mid}$ ($T_{\rm c0}$ is defined to be when the sheet resistance $R_{\rm s}$ reaches 0.1$\%$ of its normal state value $R_{\rm N}$ at $T = $3\,K and the latter is defined to be where $R_{\rm s}$ reaches 0.5$R_{\rm N}$). Inset shows the superconducting transitions in a logarithmic scale of $R(T)$ measured with the smallest current in practice, $I$ = 10 nA. $\Delta T_{\rm c}$ remains finite at the $T$ = 0 limit.
	}
	\label{Fig2} 
\end{figure*}

\begin{figure*}[htbp!]
	\centering
	\includegraphics[width=1\textwidth]{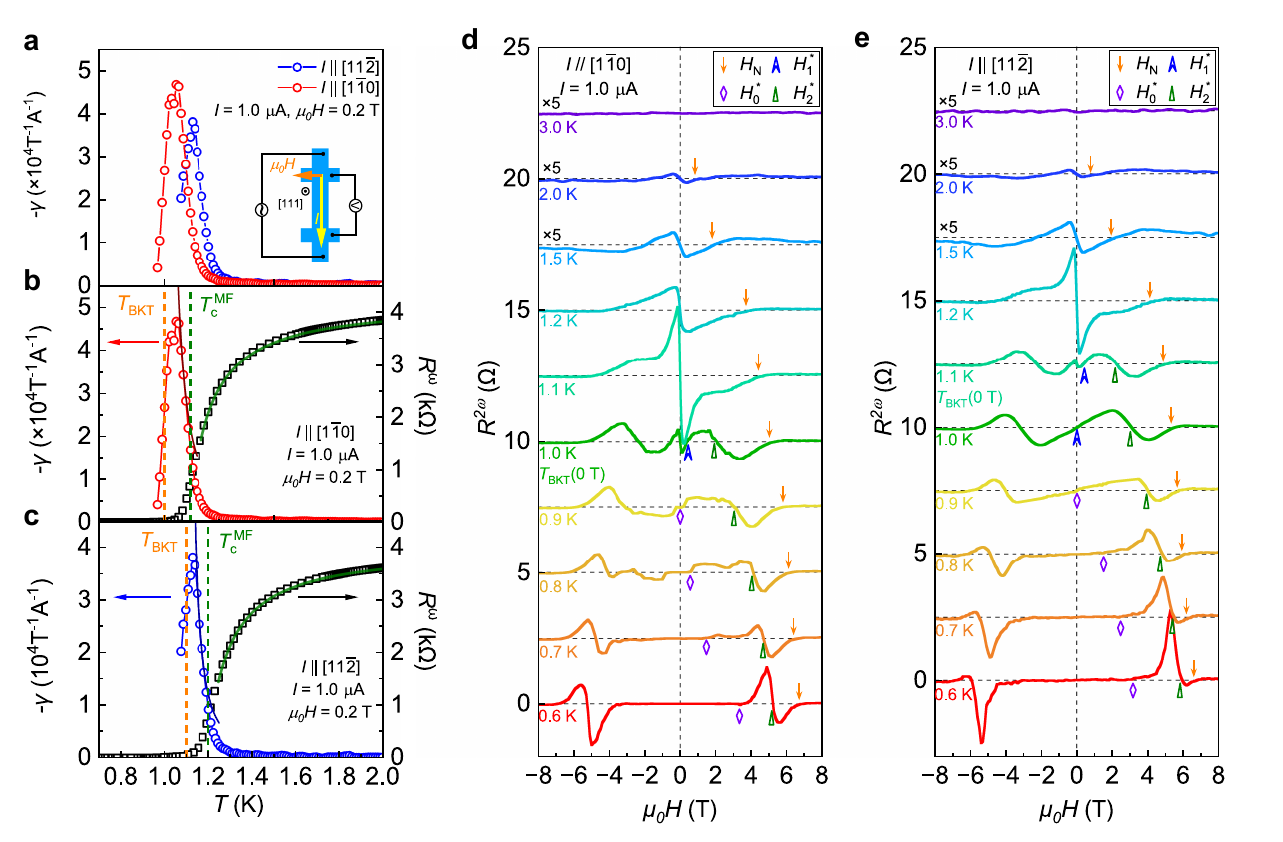}
	\caption{\textbf{Nonreciprocal charge transport in the superconducting fluctuation regime} \textbf{a}, $\gamma = \frac{2R^{2\omega}}{\mu_0 HI R^{\omega}}$ as a function of $T$ (plotted with a sign reversal). Here $\gamma$ is a coefficient characterizing the current rectification, $R^{\omega}$ and $R^{2\omega}$ are linear resistance and second harmonic resistance, respectively. Data were collected under a magnetic field $\mu_0H$ = 0.2\,T with $I$ = 1\,$\mu$A. Blue (red) symbols are data taken on the [11$\bar{2}$] ([1$\bar{1}$0]) channel of Device 3 (Supplementary Fig.\,9a, $n_{\rm 2D} = 5.59\times10^{13}$ cm$^{-2}$). Inset is a sketch of the measurement geometry for the second-harmonic nonreciprocal signal $R^{2\omega}$. \textbf{b},\textbf{c}, Comparisons of the $T$ dependence of $\gamma$ and $R^{\omega}$ measured along the [1$\bar{1}$0] direction (\textbf{b}) and the [11$\bar{2}$] direction (\textbf{c}), respectively. Approaching $T_{\rm BKT}$ (BKT transition temperature), $\gamma(T)$ is traced by a divergence $(T-T_{\rm BKT})^{-3/2}$ for both directions, which reinforces the identification of a directional-dependent $T_{\rm BKT}$ (orange vertical dashed lines). Green solid lines are fits of $R(T)$ to the Aslamazov-Lakin model (Supplementary Note 4); resulted $T_{\rm c}^{\rm MF}$ (mean-field critical temperature) are marked by olive vertical dashed lines. \textbf{d},\textbf{e}, $H$ dependence of $R^{2\omega}$ at varying $T$ for $I \parallel$ [1$\bar{1}$0] (\textbf{d}) and $I \parallel$ [11$\bar{2}$] (\textbf{e}), respectively. $R^{2\omega}(\mu_0H)$ curves are vertically shifted and data measured above 1.5\,K are re-scaled for clearance. Symbols denote characteristic magnetic fields: $H_0^*$ (purple diamonds) and $H_N$ (orange arrows) are the onset and termination of the finite $R^{2\omega}$ due to superconductivity~\cite{WakatsukiMoS2}, respectively; $H_1^*$ (blue arrowheads) and $H_2^*$ (green up triangles) denote two characteristic fields at which $R^{2\omega}(H)$ changes its sign.
	}
	\label{Fig3}
\end{figure*}

\begin{figure*}[htbp!]
	\centering
	\includegraphics[width=0.85\textwidth]{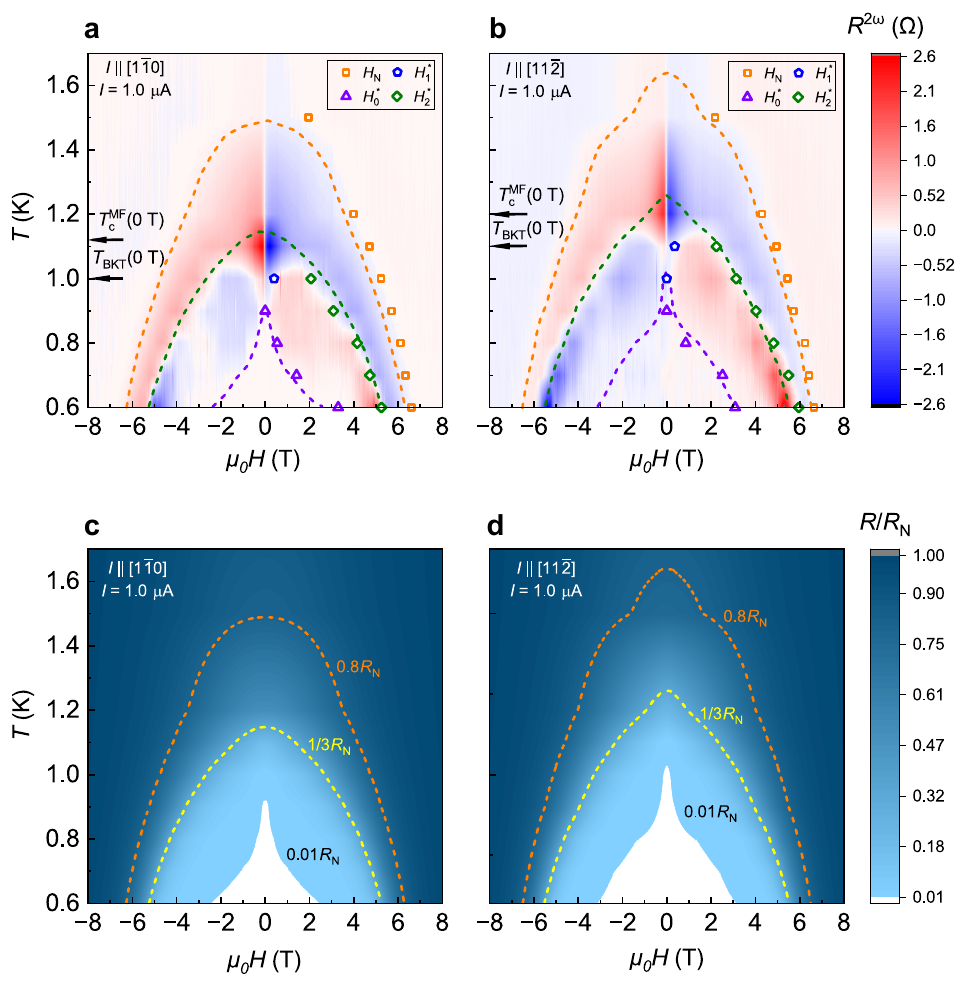}
	\caption{\textbf{Directional-dependent contour plots $R^{2\omega}$ and $R^{\omega}$ in the $H-T$ plane.} \textbf{a},\textbf{b}, Color contour plots of $R^{2\omega}(T,H)$ for $I$ along [1$\bar{1}$0] (\textbf{a}) and [11$\bar{2}$] (\textbf{b}), respectively, constructed based on the data shown in Figs.\,3d and e. $T_{\rm BKT}$ and $T_{\rm c}^{\rm MF}$ (BKT transition temperature and mean-field critical temperature) for each direction are denoted by black arrows on the vertical axis. $R^{\omega}$ and $R^{2\omega}$ are linear resistance and second harmonic resistance, respectively. $H_0^*$ (purple triangles), $H_1^*$ (blue pentagons), $H_2^*$ (green diamonds) and $H_N$ (orange squares) are represented by symbols overlaid to the contour plots. \textbf{c},\textbf{d}, Contour maps of the normalized magnetoresistance $\Delta R(H)/R_{\rm N}$ [$\Delta R(H) = R(H)-R(H =0)$] for $I \parallel$ [1$\bar{1}$0] (\textbf{c}) and $I \parallel$ [11$\bar{2}$] (\textbf{d}), respectively (see Supplementary Fig.\,9 for raw data). Orange and yellow short-dashed lines are contour lines of $R$ = 0.8\,$R_{\rm N}$ and (1/3)\,$R_{\rm N}$, respectively. The boundary of the white area denotes the contour line of $R$ = 0.01\,$R_{\rm N}$ . These three contour lines are reproduced in \textbf{a} and \textbf{b} to underline their correspondence with the characteristic fields.
	}
	\label{Fig4}
\end{figure*}

\newpage

\begin{center}
{\large Supplementary Information for} \\
{\large \bf Directional-dependent Berezinskii-Kosterlitz-Thouless transition at EuO/KTaO$_3$(111) interfaces}

\bigskip

Zongyao Huang$^1$,
Zhengjie Wang$^1$, Xiangyu Hua$^2$, Huiyu Wang$^1$, Zhaohang Li$^1$, Shihao Liu$^1$, Zhiwei Wang$^1$, Feixiong Quan$^1$,
Zhen Wang$^1$, Jing Tao$^1$, James Jun He$^{3}$,
Ziji Xiang$^{2,3}$
Xianhui Chen$^{1,2,3}$

\medskip
{\it \small
\setlength{\baselineskip}{16pt}

    $^1$ Department of Physics, University of Science and Technology of China, Hefei 230026, China\\
	$^2$ Hefei National Research Center for Physical Sciences at the Microscale, University of Science and Technology of China, Hefei 230026, China\\
	$^3$ Hefei National Laboratory, University of Science and Technology of China, Hefei 230088, China.
}

\end{center}
\bigskip



\newpage

\begin{table*}[htb]
\renewcommand{\thetable}{S1}
\centering
\caption{\textbf{Growth conditions of EuO/KTaO$_{3}$(111) interface samples.} Growth temperature $T_{\rm growth}$ and oxygen partial pressure $p(\rm O_2)$ for seven samples with different 2D carrier density $n_{\rm 2D}$ and zero-resistance temperature $T_{\rm c0}$ (along two orthogonal directions, [11$\bar{2}$] and [1$\bar{1}$0], respectively). Samples are listed in the order of (low to high) $n_{\rm 2D}$. Numbering of the devices patterned on corresponding samples is also noted. Here the values of $T_{\rm c0}$ (displayed in main text Fig.\,1b) were obtained from measurements using the van der Pauw configuration and hence can be slightly different from those determined from fabricated Hall-bar devices.}
\label{tab1}
\begin{ruledtabular}
\begin{tabular}{c|c|ccccc}

  Sample No. & Device No. & $n_{\rm 2D}$ ($10^{13}$ cm$^{-2}$)  & $T_{\rm c0}[11\bar{2}]$ (K) & $T_{\rm c0}[1\bar{1}0]$ (K) & $T_{\rm growth}$ ($^\circ$C) & $p(\rm O_2)$ (10$^{-9}$ mBar) \\\hline
  2 & 2 & 5.24 & 1.23 & 1.15 & 450 & 2.2 \\
  3 & 3 & 5.59 & 1.12 & 0.88 & 450 & 2.0 \\
  9 & 5 & 7.40 & 1.33 & 1.28 & 450 & 1.5 \\
  8 & 4 & 7.46 & 1.36 & 1.32 & 450 & 1.5 \\
  1 & 1 & 7.67 & 1.22 & 1.07 & 450 & 1.5\\
  4 & & 7.90 & 1.43 & 1.34 & 450 & 1.3  \\
  5 & & 8.19 & 1.48 & 1.31 & 450 & 1.0  \\
  6 & & 8.34 & 1.63 & 1.58 & 450 & 1.0  \\
  10 & 6 & 8.80 & 1.60 & 1.60 & 500 & 1.0 \\
  7 & & 9.79 & 1.70 & 1.70 & 500 & 0.5  \\

\end{tabular}
\end{ruledtabular}
\end{table*}
\bigskip

\noindent
\textbf{Note 1: Interface morphology of the EuO/KTaO$_3$(111) heterostructure}

\noindent
To achieve direct imaging of the interface, scanning transmission electron microscopy (STEM) measurements have been conducted on a Thermo Scientific Themis Z microscope equipped with a probe-forming spherical aberration corrector. Data were taken on Sample No.\,2 in two high-symmetry crystallographic planes of the single-crystalline KTaO$_3$ (KTO) substrate: (1$\bar{1}$0) and (11$\bar{2}$), as shown in Supplementary Figs.\,2b and c, respectively. The sample was prepared using focused ion beam (Gallium) in advance. A relatively abrupt interface was observed, in agreement with previous reports of EuO/KTO interfaces \cite{ChangjiangLiu1SI,HuaSOCSI}; the interfacial disorder, manifested as lattice distortion, oxygen vacancy and atomic intermixing (most notably the substitution of Eu on K sites), occurs only within the topmost 3 unit cells of KTO. This observation thus effectively delineates an interface region with a width $\simeq$ 1\,nm on the KTO side.

We note that the EuO overlayer is polycrystalline (Supplementary Figs.\,2d and e). Nevertheless, KTO is still in a crystalline state near the interface with clearly defined crystallographic directions, as demonstrated in Supplementary Figs.\,2b and c (note the atomic configurations consistent with the expectation shown in Supplementary Fig.\,2a). Moreover, the conducting layer in KTO (with a typical thickness of 5-8\,nm for KTO-based interfaces \cite{Zhang2DEGSI,HuaSOCSI,YanwuXieScience,YanwuXiePRLSI}) is appreciably thicker than the disordered interface region ($\simeq$ 1\,nm). Note that the superconducting layer thickness is estimated to be 7-9\,nm for our samples (Supplementary Table 2). As such, we point out that the interfacial two-dimensional electron gas (2DEG) residing in the KTO surface layers \cite{Zhang2DEGSI} retains the well-defined energy-momentum dispersion relations inherited from crystalline bulk KTO; the major part of 2DEG occurs in a depth that is generally free of interfacial element substitution or lattice distortion. Hence, the crystalline azimuthal dependence of physical properties likely reflects the intrinsic behaviour of 2DEG, and our analysis based on symmetry arguments is validated.

\bigskip

\noindent
\textbf{Note 2: Magnetic properties of the interface samples}

\noindent
The polycrystalline EuO overlayers exhibit almost identical magnetic properties with the (111)-oriented EuO epitaxial films [grown on KTO(110) surfaces] reported in Ref.\cite{HuaStripeSI}. As illustrated in Supplementary Figs.3a and b, magnetization ($M$) measurements taken in a superconducting quantum interference device (SQUID) reveals a ferromagnetic transition occurring at the Curie temperature $T_{\rm Curie} \simeq$ 73\,K; saturation fields determined from the $M(H)$ isotherms (Supplementary Fig.\,3a) appear to be higher with an out-of-plane magnetic field ($H \parallel$ [111]), corroborating an easy-plane magnetic anisotropy of the EuO film. Considering that the magnetization of EuO contributes remarkably to the total magnetic flux density $B = \mu_0(H+M)$ ($\mu_0$ is the vacuum permeability), we plot all the physical quantities against the external $H$-field in this study (the $M$ term gives a correction up to 1.38\,T in the $B$-field).

The correspondence between directional-dependent $T_{\rm c}$ and signatures of ferromagnetism in normal-state transport properties [e.g., anomalous Hall effect (AHE) and hysteresis loops in magnetoresistance (MR)] has been underlined for EuO/KTO(110) 2DEGs \cite{HuaStripeSI}. Here, we stress that such correspondence also manifests itself in EuO/KTO(111) samples. As displayed in Supplementary Fig.\,3c, a non-zero anomalous Hall resistance $R_{\rm xy}^{\rm A}$ can be resolved in all samples saving Sample No.\,7 (with the highest $n_{\rm 2D}$ = 9.79$\times$10$^{13}$ cm$^{-2}$) after subtracting the $H$-linear ordinary Hall. All 2DEGs exhibiting AHE are characterized by different $T_{\rm c}$ along [11$\bar{2}$] and [1$\bar{1}$0] directions, and vice versa (Supplementary Fig.\,4, results summarized in main text Fig.\,1b). The $H$ dependence of $R_{\rm xy}^{\rm A}$ emulates that of $M$ with $H \parallel $ [111] (Supplementary Fig.\,3a): Both rises monotonically with increasing $H$ before saturating at $\mu_0H \approx$ 2.8\,T. The resemblance between the profiles of $R_{\rm xy}^{\rm A}(H)$ and $M(H)$, together with the concurrence of AHE and ferromagnetic order at $T_{\rm Curie}$ (inset of Supplementary Fig.\,3b), points towards the proximity-induced ferromagnetism \cite{Zhang2DEGSI} for the interfacial 2DEGs. That is, the ferromagnetic order of 2DEG stems from the spin polarization effect triggered by the ferromagnetically-ordered Eu 4$f$ moments in the EuO overlayer, instead of Stoner-type instabilities of the itinerant Ta 5$d$ electrons themselves \cite{CaZrO3StonerSI} or localized 5$d$ electrons engendered by oxygen vacancies \cite{WeiLiFMSI}.  As pointed out by previous band calculations for a EuO/KTO supercell \cite{HuaStripeSI}, the $d-f$ hybridization effect (which promotes the spin polarization of the conduction electrons) is strongest near the bottom of Ta 5$d$ band; upon increase of band filling, the chemical potential moves up and the spin splitting of Fermi surface becomes promptly reduced. This picture is in accord with the observation that the AHE is missing in the sample with the highest $n_{\rm 2D}$. On the other hand, a clear relationship between the amplitude of $R_{\rm xy}^{\rm A}$ and $n_{\rm 2D}$ cannot be established at this stage. Features of ferromagnetic 2DEG are further accentuated by the MR data: With $H$ applied in-plane, a bow-tie-shaped hysteretic MR loop is observed at low fields for Sample No.\,1 ($n_{\rm 2D}$ = 7.67$\times$10$^{13}$) cm$^{-2}$, but is absent in Sample No.\,7 (Supplementary Fig.\,3d). We conclude that the directional superconductivity ubiquitously emerges from ferromagnetic 2DEGs at the EuO/KTO interfaces.

\bigskip
\noindent
\textbf{Note 3: Upper critical fields and 2D superconductivity}
\par

\begin{table*}[htb]
\renewcommand{\thetable}{S2}
\centering
\caption{\textbf{Directional-dependent Ginzburg-Landau coherence lengths.}
The zero-$T$ GL coherence length $\xi_{\rm GL}$(0) and superconducting layer thickness $d$ calculated based on Eqs.\,\ref{Hc2perp} and \ref{Hc2para} for Device 1.
$\xi_{\rm GL}^{[11\bar{2}]}$(0) and $\xi_{\rm GL}^{[1\bar{1}0]}$(0) denote the coherence lengths along the $[11\bar{2}]$ and $[1\bar{1}0]$ directions, respectively. $d < \xi_{\rm GL}$ corroborates the 2D nature of interface superconductivity.}
\label{tab2}
\par
\begin{ruledtabular}
\begin{tabular}{c|c|c|c}

   & $\xi_{\rm GL}^{[11\bar{2}]}$(0) (nm) & $\xi_{\rm GL}^{[1\bar{1}0]}$(0) (nm) & $d$ (nm) \\\hline
  $I \parallel [11\bar{2}]$ & 19.9(3) & 17.4(2) & 7.8(1) \\\hline
  $I \parallel [1\bar{1}0]$ & 22.2(4) & 19.6(3) & 8.8(1)

\end{tabular}
\end{ruledtabular}
\end{table*}

\noindent
The information of the anisotropic upper critical field ($H_{\rm c2}$) was collected from the temperature-dependent resistance ($R(T)$) curves measured on Hall-bar Device 1 (inset of main text Fig.\,1c) under magnetic fields with varying strengths and orientations. The raw data are presented in Supplementary Figs.5a-f. For each current direction ($I \parallel$ [11$\bar{2}$] and $I \parallel$ [1$\bar{1}$0]), we measured the $R(T)$ data with three orthogonal orientations of $H$: [111] (out-of-plane), [11$\bar{2}$] and [1$\bar{1}$0] (both are in-plane). At a given $H$, the transition temperature $T_{\rm c}$ in this subsection is defined as the temperature where $R$ reaches half of the normal state value $R$($T$ = 3\,K); correspondingly, the $T_{\rm c}(H)$ plot translates to $H_{\rm c2}(T)$ by switching the coordinate axes. According to the Tinkham's formula based on 2D Ginzburg-Landau (GL) theory \cite{Tinkham,2DHc}, the out-of-plane and in-plane $H_{\rm c2}$ can be expressed as:

\begin{equation}
\mu_0H_{\rm c2}^{\perp}(T) = \frac{\Phi_0}{2\pi \xi_{\rm GL}^2(0)}(1-\frac{T}{T_c});
\label{Hc2perp}
\end{equation}

\begin{equation}
\mu_0H_{\rm c2}^{\parallel}(T) = \frac{\Phi_0 \sqrt{12}}{2\pi d\xi_{\rm GL}(0)}\sqrt{1-\frac{T}{T_c}}.
\label{Hc2para}
\end{equation}

\noindent
Here $\Phi_0$ = $h/(2e)$ is the magnetic flux quantum ($h$ is the Planck constant), $\xi_{\rm GL}$(0) is the GL coherence length at zero temperature and $d$ is the thickness of superconducting layer. As depicted in Supplementary Figs.\,5g-i, the $T$ dependence of $H_{\rm c2}^{\perp}$ and $H_{\rm c2}^{\parallel}$ measured on Device 1 can be satisfyingly fitted by Eqs.\,\ref{Hc2perp} and \ref{Hc2para} (dashed lines), respectively, for both current directions. The fits yield $d \simeq$ 8-9\,nm and a coherence length $\xi_{\rm GL}$(0) that depends on both the orientation of $H$ (i.e., showing in-plane anisotropy) and the direction of current $I$ (i.e., showing directional dependence), as summarized in Supplementary Table\,2. $H_{\rm c2}$ (and hence $\xi_{\rm GL}$) exhibits a moderate in-plane anisotropy, manifesting as a larger $\xi_{\rm GL}$ along [11$\bar{2}$] exceeding that along [1$\bar{1}$0] by 11-12$\%$; such anisotropy implicitly reflects the sixfold rotational symmetry of the 2D Fermi surfaces \cite{UPt3sixfold} on the KTO(111) surface. Moreover, the directional dependence of $\xi_{\rm GL}$ is comparable with the in-plane anisotropy in amplitude; the channel with higher $T_{\rm c}$ (the [11$\bar{2}$] channel) consistently exhibits higher $H_{\rm c2}$ for all three $H$ orientations, highlighting the directional nature of the interface superconductivity. These phenomena have also been observed for the EuO/KTO(110) interfaces \cite{HuaStripeSI}. We further note that the zero-temperature values of $H_{\rm c2}^{\parallel}$ are remarkably higher than the Pauli paramagnetic limit (Supplementary Figs.\,5h and i), which may imply the profound influence of spin-orbit scattering at the interface \cite{HuaSOCSI}.

\bigskip

\noindent
\textbf{Note 4: Analysis of paraconductivity and superconducting fluctuations}

\noindent
The paraconductivity, i.e., the contribution of superconducting fluctuations to the conductivity in zero $H$, can be analyzed separately in different regimes of $T$. The well-known Aslamazov-Larkin (AL) paraconductivity \cite{ALoriginal} describes the effect of Gaussian-type fluctuation of the superconducting order parameter, which involves short-lived Cooper pairs emerging above the mean-field critical temperature $T_{\rm c}^{\rm MF}$. The resulted power-law divergence of superconducting correlation length as $T$ is reduced towards $T_{\rm c}^{\rm MF}$ delineates the expression of AL paraconductivity. In the 2D case and for $T$ not very far from $T_{\rm c}^{\rm MF}$, the expression is given as \cite{ALoriginal,ALCaprara,BenfattoBroadening}:
\begin{equation}
\Delta G_{\rm AL}= \frac{e^2}{16\hbar}\frac{T_{\rm c}^{\rm MF}}{T-T_{\rm c}^{\rm MF}},
\label{2DAL}
\end{equation}

\noindent
where $\Delta G(T)$ = $G(T) - G_{\rm N}(T)$ = $1/R(T)-1/R_{\rm N}$ represents the excess conductance originating from (Gaussian) superconducting fluctuations, $R_{\rm N}$ is the nomal-state resistance of the 2DEG that is assumed to be $T$-independent below 3\,K (that is, $R_{\rm N}$ = const. in the range of $T$ of interest), $\hbar$ is the reduced Planck constant. Consequently, we have:

\begin{equation}
\frac{R(T)}{R_{\rm N}} = (1+\frac{e^2R_{\rm N}}{16\hbar}\frac{T_{\rm c}^{\rm MF}}{T-T_{\rm c}^{\rm MF}})^{-1}.
\label{ALfit}
\end{equation}
\noindent
Eq.\,\ref{ALfit} is the fitting formula used for analyzing the AL paraconductivity in this work. As illustrated in Supplementary Fig.\,6, the AL model denoted by black solid curves fits the experimental $R(T)$ curves well in the selected ranges of $T$. The AL fits provide a reasonable estimation of $T_{\rm c}^{\rm MF}$ for each Hall-bar channel (see Supplementary Table 3); the values of $T_{\rm c}^{\rm MF}$ mentioned in the main text were attained by this method. Note that the AL fitting results of $T_{\rm c}^{\rm MF}$ are consistent with those determined from the inflection point of $R(T)$ (main text Figs.\,1d and e, insets), which serves as an alternative criterion for defining $T_{\rm c}^{\rm MF}$ \cite{BenfattoLa214SI}.

\begin{table*}[htb]
\renewcommand{\thetable}{S3}
\renewcommand*{\thetable}{S\arabic{table}}
\centering
\caption{\textbf{Characteristic parameters attained from analysis of paraconductivity.} Directional-dependent parameters extracted from fits of $R_{\rm s}(T)$ measured on different channels of Devices 1-3. The mean-field critical temperature $T_{\rm c}^{\rm MF}$ and the BKT transition temperature $T_{\rm BKT}$ are determined from Eq.\,\ref{ALfit} and Eq.\,\ref{HN}, respectively. The separation between $T_{\rm c}^{\rm MF}$ and $T_{\rm BKT}$ is reflected by the parameter $t_{\rm c}=(T_{\rm c}^{\rm MF} - T_{\rm BKT})/T_{\rm BKT}$. $t_{\rm c}$ and the coefficient $b$ in Eq.\,\ref{HN} provide an estimation of $\mu/J_{\rm s}$: $b = (4/\pi^2)(\mu/J_{\rm s})\sqrt{t_{\rm c}}$. }
\label{tab3}
\begin{ruledtabular}
\begin{tabular}{c|c|c|c|c|c|c}

  & & $T_{\rm c}^{\rm MF}$ (K) & $T_{\rm BKT}$ (K) & $b$ & $t_{\rm c}$ & $\mu/J_{\rm s}$  \\\hline
Device 1 & $[11\bar{2}]$ channel            & 1.251(3)  & 1.230(3)  & 0.23(3) & 0.017 & 4.4(7)  \\\hline
         & $[1\bar{1}0]$ channel            & 1.169(4)  & 1.149(2)  & 0.24(3) & 0.017 & 4.5(8)  \\\hline
Device 2 & $[11\bar{2}]$ channel(cyan)      & 1.482(1)  & 1.412(9)  & 0.30(4) & 0.049 & 3.4(6)  \\\hline
         & $[11\bar{2}]$ channel(green)     & 1.321(1)  & 1.258(7)  & 0.37(5) & 0.050 & 4.1(6)  \\\hline
         & $[11\bar{2}]$ channel(navy)    & 1.302(1)  & 1.232 8)  & 0.39(6) & 0.057 & 4.0(7)  \\\hline
Device 2 & $[1\bar{1}0]$ channel(brown)  & 1.282(9)  & 1.203(4)  & 0.56(2) & 0.066 & 5.4(4)  \\\hline
         & $[1\bar{1}0]$ channel(amber gold)    & 1.194(1)  & 1.164(4)  & 0.27(4) & 0.026 & 4.2(6)  \\\hline
         & $[1\bar{1}0]$ channel(red)       & 1.184(6)  & 1.140(7)  & 0.32(5) & 0.039 & 4.0(7)  \\\hline
Device 3 & $[11\bar{2}]$ channel            & 1.199(1)  & 1.112(1)  & 0.63(1) & 0.080 & 5.5(1)  \\\hline
         & $[1\bar{1}0]$ channel            & 1.117(1)  & 1.023(2)  & 0.62(1) & 0.095 & 5.0(1)
\end{tabular}
\end{ruledtabular}
\end{table*}

Below $T_{\rm c}^{\rm MF}$, the AL paraconductivity no longer exists, yet the vortex-driven phase fluctuations become significant. In the BKT scenario, these are associated with the thermally-activated vortexlike topological excitations above the transition temperature $T_{\rm BKT}$, which proliferate upon warming and result in a correction to the 2D conductance. This term of paraconductivity is commonly described by the Halperin-Nelson (HN) formula \cite{HNformulaSI}:
\begin{equation}
\frac{R(T)}{R_{\rm N}}=[1+\frac{4}{A^2}\sinh^2(\frac{b}{\sqrt{t}})]^{-1},
 \label{HN}
\end{equation}
\noindent
where $t$ = $(T-T_{\rm BKT})/T_{\rm BKT}$, $A$ is a dimensionless constant of order 1 and $b \simeq 2a\sqrt{t_c}$ is a constant related to some characteristic properties of the system: $t_c= (T_{\rm c}^{\rm MF}-T_{\rm BKT})/ T_{\rm BKT}$, $a = \mu(T)/\mu_{\rm XY}$, $\mu(T)$ is the vortex core energy and $\mu_{\rm XY}$ is the value of $\mu$ in the 2D XY model \cite{BenfattoBroadening,BenfattoLa214SI}. The 2D XY model also predicts a constant ratio between $\mu_{\rm XY}$ and the superfluid stiffness $J_{\rm s}$: $\mu_{\rm XY} \simeq (\pi^2/2) J_{\rm s}$ \cite{Benfattolayered,NbN}. The fitting results of the HN formula are summarized in Supplementary Table 3. In all of our devices, the intervals between $T_{\rm BKT}$ and $T_{\rm c}^{\rm MF}$ are small: $t_c \simeq 1-10\%$; the proximity between $T_{\rm BKT}$ and $T_{\rm c}^{\rm MF}$ has been discussed in other oxide interfaces \cite{BenfattoBroadening,LiuNatComm} as well as thin films of conventional superconductors \cite{NbN}. According to the Beasley-Mooij-Orlando approximate relation \cite{BMOSI},
\begin{equation}
\frac{T_{\rm BKT}}{T_{\rm c}^{\rm MF}} \approx (1+0.173\times\frac{R_{\rm N}}{R_C})^{-1},
 \label{BMO}
\end{equation}
\noindent
here $R_C = \hbar/e^2 \approx$ 4.12\,k$\Omega$. The normal-state sheet resistance $R_{\rm N} \approx 400-1400\,\Omega$ yields $T_{\rm BKT}$/$T_{\rm c}^{\rm MF} \simeq $ 0.94-0.99, qualitatively consistent with the fitting results shown in Supplementary Table 3.

The disparity of $T_{\rm BKT}$ between the $[11\bar{2}]$ and $[1\bar{1}0]$ channels can be further elucidated based on the HN analysis. In principle, such a disparity can be a consequence of the anisotropic vortex core energy $\mu$ (and hence the parameter $b$ in Eq.\,\ref{HN}); when $\mu$ profoundly deviates from $\mu_{\rm XY}$ ($a \neq$ 1), the superconducting transition takes place at a temperature either higher (with $a \gg 1$) \cite{Benfattolayered} or lower ($a \ll$ 1) \cite{NbN} than that predicted by the 2D XY model, i.e., application of main text Eq.\,1 to the Bardeen-Cooper-Schrieffer (BCS) profile of $J_{\rm s}(T)$. Nevertheless, we determined that this is not the case in our devices: The fitted values of $b$ show no distinct dependence on the current direction (this is clearly evidenced by the comparison among channels of Device 2), suggesting the absence of in-plane anisotropy for $\mu$ (Supplementary Table\,3). The directional-dependent $T_{\rm BKT}$ must has an origin different from the anisotropic vortex core energy. A ratio $\mu/J_{\rm s} \simeq $3.4-5.5 can be deduced from the HN fitting results of $a = \mu/\mu_{\rm XY} \simeq$ 0.67-1.11, which is larger than that in NbN films ($\mu/J_{\rm s} \simeq $0.5-2.2, \cite{NbN,YongNbNSI}) but is smaller than that in films of high-$T_{\rm c}$ cuprates ($\mu/J_{\rm s} >$ 6, \cite{BenfattoLa214SI,BenfattoBroadening}).

Most of our devices exhibit a small resistive tail below $T_{\rm BKT}$ (Supplementary Fig.\,6). It can be due to two reasons: Finite-size effect and inhomogeneity. An earlier analysis points out that the finite-size effect is less pronounced in systems with small $t_c$ (because in such case the divergent correlation length $\xi$ remains appreciably large below $T_{\rm BKT}$) \cite{BenfattoBroadening}, which is applied to the present instance. In this context, we attribute the BKT tail behavior mainly to the inherent inhomogeneity at the oxide interfaces \cite{BenfattoBroadening,CapraraSI,VendittiIVSI}. An inhomogeneous interface is usually depicted as an ensemble of randomly-distributed superconducting patches, each hosting an individual local $T_{\rm BKT}$. Nevertheless, recent measurement of $J_{\rm s}$ reveals a very narrow Gaussian distribution of $T_{\rm BKT}$ at the KTO-based interfaces \cite{KTOsuperfluidSI}, validating the global BKT description of these samples (in contrast, SrTiO$_3$-based interfaces are characterized by stronger inhomogeneity \cite{CapraraSI} and the BKT-like behavior of $J_{\rm s}$ has never been established \cite{STOsuperfluid,Bimodal}). We also emphasize that a random local critical temperature $T_{\rm BKT}$ is totally irrelevant to directional dependence we observed, since the later clearly clings to the KTO crystal axes.

\bigskip

\noindent
\textbf{Note 5: Measurement of $I-V$ characteristics}
\par
\noindent
We performed tests of the $I-V$ relationships on two Hall-bar channels of Device 2 (see Supplementary Fig.\,8a): The $[11\bar{2}]$ channel with the highest $T_{\rm BKT}$ (marked by cyan color in Supplementary Fig.\,8a) and the $[1\bar{1}0]$ channel with the lowest $T_{\rm BKT}$ (marked by red color); these two are perpendicular to each other. Considering the large sample resistance (6-9 k$\Omega$) and the limited cooling power of the dilution refrigerator insert, we adopted a sweeping mode to avoid severe Joule heating and the resulted temperature drift. The current $I$ was swept at a rate of approximately 25 nA/s. The system temperature was monitored by a thermometer attached to the sample mount and we ensured no drift in its reading throughout the measurement process. The collected $I-V$ data are presented in Supplementary Figs.8c-f; the data of $\alpha(T)$ presented in main text Fig.\,2c were extracted from the slope of $I-V$ curves (in the log-log plots, Supplementary Figs.\,8c and d). We mention that the nonlinear $I-V$ characteristics, hallmark of the BKT physics, occurs only beyond a low-current ``Ohmic" regime that stems from the Lorentz-force-induced breaking of large (and weak) vortex-antivortex bound pairs in a finite-size system \cite{ReyrenSI,fluxflow}. As shown in Supplementary Figs.\,8e and f, the $V(I)$ curves measured at lower $T$ display a series of step-like jumps; such signatures disappear when $T$ reaches about $T_{\rm BKT}$-0.2\,K. These voltage jumps can be invoked via two mechanisms: (i) The hot spots \cite{hotspot}, i.e., certain regions of the sample are heated to the normal state owing to the inhomogeneous current dissipation in a disordered system; (ii) the Larkin-Ovchinnikov-type flux-flow instability \cite{LOinstability}, which describes the unstable $I-V$ characteristics created by  ultra-fast moving vortices. Under forward and reverse current bias sweeps, (i) yields an anticlockwise hysteresis of the $I-V$ curves, whereas (ii) may promote both anticlockwise and clockwise (in the presence of strong inelastic quasiparticle scattering \cite{fluxflow,Mo3Si}) hysteresis loops. The narrow clockwise hysteresis observed in our data (Supplementary Figs\,.8e and f) points towards the latter case, highlighting the relevance of fast-moving vortices at the interface.  It should also be mentioned that Josephson-like dynamics has been revealed at LaAlO$_3$/SrTiO$_3$ interfaces \cite{JosephsonIV1,JosephsonIV2}; in such cases, the superconducting state is described as a random array of Josephson-coupled superconducting islands, the supercurrent is thus established by tunneling of Cooper pairs across weak links between these islands. Such scenario implicitly suggests strong electron inhomogeneity at the interface. Nevertheless, the most prominent feature of Josephson-like dynamics, i.e., pronounced anticlockwise hysteresis in the $I-V$ curves down to the lowest $T$ \cite{JosephsonIV1,JosephsonIV2}, is absent at our KTO-based interfaces, thus defying the Josephson-junction array description.

It has been established that main text Eq.\,1 is valid even with the effect of vortex core energy $\mu \neq \mu_{\rm XY}$ taken into account (whereas $J_{\rm s}(T)$ deviates from the BCS profile at a $\mu$-dependent $T < T_{\rm BKT}$) \cite{BenfattoLa214SI,slowmotion}. Therefore, the nonlinear relationship $V \propto I^{\alpha}$, $\alpha = 1+\pi J_{\rm s}/k_B T$ leads to the expectation of $\alpha(T_{\rm BKT})$ = 3 in spite of the various values of $\mu$ in our samples. Nonetheless, the combined influence of $\mu$ and inhomogeneity naturally causes a smeared BKT transition \cite{BenfattoBroadening,VendittiIVSI} without the sharp jump of $\alpha$ from 3 to 1 expected for the ideal case. We note that in Device 2, the linearity of $I-V$ characteristics ($\alpha$ = 1) is recovered at $T \simeq 1.2-1.3\,T_{\rm BKT}$ (main text Fig.\,2c); this contrasts with the nonlinear $I-V$ characteristics of SrTiO$_3$-based interfaces, wherein $\alpha$ = 1 only occurs at an abnormally high $T \simeq 1.8-2.0\,T_{\rm c0}$ ($\gtrsim$ 1.5 times the ``apparent" $T_{\rm BKT}$ determined by $\alpha(T_{\rm BKT})$ = 3) \cite{VendittiIVSI}. The latter case point towards a failure of the BKT description and the superconducting transition is instead established through a percolation process \cite{percolation}. The weaker inhomogeneity at our KTO-based interfaces compared with their SrTiO$_3$-based counterparts is consistent with the results of superfluid stiffness measurements as mentioned above \cite{KTOsuperfluidSI}.

\bigskip

\noindent
\textbf{Note 6: Nonreciprocal charge transport}
\par
\noindent
\underline{A. The temperature dependence}
\par
\noindent
It has been well established that a system in which both the time reversal symmetry (TRS) and the inversion symmetry (IS) are broken exhibits nonreciprocal charge transport \cite{BiTeBr,Nonreciprocal}, i.e., the forward and backward current flows trigger different voltage responses. For the oxide heterointerfaces, the inherent IS breaking owing to the structural asymmetry endows the interfacial 2DEGs with Rashba-type spin-orbit interaction; under the application of an in-plane $H$-field, a nonreciprocity taking the form $\Delta R = R(I)-R(-I) \propto$ \textbf{I}$\cdot$(\textbf{P}$\times$\textbf{B}), where \textbf{P} is the polarization vector that points along the normal direction of the interface \cite{ChoeLAO-STOSI}, $B$ is the magnetic flux density. This nonreciprocity is more conveniently captured by the measurement of second harmonic signal $R^{2\omega}$; the coefficient $\gamma$ describing the relative amplitude of nonreciprocal resistance (with respect to the normal resistance) can be inferred from the ratio of $R^{2\omega}$ to the linear term $R^{\omega}$ \cite{BiTeBr}:
\begin{equation}
\gamma = \frac{2R^{2\omega}}{R^{\omega}BI}.
 \label{secondharmonic}
\end{equation}
\noindent
$\gamma$ is a widely used parameter characterizing the effect of the current rectification in real systems; note that its definition requires the nonreciprocal response to satisfy $R^{2\omega} \propto BI$, which is valid in the low-field, low-current region. In our EuO/KTO samples, whereas the symmetry allows the presence of finite $\gamma$ in the normal state, such signal is rather weak ($\gamma \lesssim $ 200\,T$^{-1}$A$^{-1}$, which is mostly beyond the experimental resolution, see Supplementary Fig.\,9b); remarkable nonreciprocal charge transport emerges once the influence of superconducting fluctuations (paraconductivity) sets in (main text Fig.\,3). This is due to the significantly reduced energy scale $\sim k_BT_{\rm c}$ in the superconducting regime (compared with $E_F$ in the normal state, $\gamma$ is inversely proportional to the cube of such energy scale) \cite{WakatsukiMoS2SI}. The most notable feature in the $T$ dependence of $\gamma$ is the drastic enhancement when $T$ approaches $T_{\rm BKT}$, which develops a $(T-T_{\rm BKT})^{-3/2}$ divergence in a specific temperature range slightly above $T_{\rm BKT}$ (main text Figs.\,3a-c); such a power-law divergence, a hallmark of the contribution of thermally-activated vortexlike excitations preceding a BKT transition, has been repeatedly observed in 2D noncentrosymmetric superconductors \cite{HoshinoSI,ItahashiSTOSI,MasukoSI}. We emphasize that since the $H$-field is applied parallel to the superconducting interface (here we replace $B$ in Eq.\,\ref{secondharmonic} with $\mu_0H$ for clarity), field-induced vortices are irrelevant in our experiments and the nonreciprocal vortex dynamics only refers to the thermally created vortices (and antivortices) in the BKT theory. The divergent $\gamma$ near the BKT transition stems from the renormalization of superfluid density \cite{HoshinoSI}. Above $T_{\rm c}^{\rm MF}$, the vortex-induced rectification effect vanishes as the phase fluctuations cease to play a role in paraconductivity; in this regime, the nonreciprocal paraconductivity (see the residual ``tail" of $\gamma$ extending to $T > T_{\rm c}^{\rm MF}$, main text Figs.\,3b and c) originates from the Gaussian-type (amplitude) fluctuations described by the AL model \cite{HoshinoSI,ItahashiSTOSI}. Both the Gaussian fluctuation and the vortex flow mechanism yield a nonreciprocal signal with a negative sign, i.e., $\gamma <$ 0.  In Supplementary Fig.\,10d, we plot the current dependence of $R^{2\omega}$, which shows linearity in the low-current region (e.g., $|I| \lesssim $3\,$\mu$A at 1.7\,K), yet develops a maximum and eventually becomes suppressed upon further increase of $|I|$. Such behavior is qualitatively in accord with the earlier report in polar superconductor SrTiO$_3$ \cite{ItahashiSTOSI}. We also mention that a current amplitude of 1\,$\mu$A selected for the majority of $R^{2\omega}$ measurements (see data in main text Figs.\,3 and 4) satisfies the linear response requirement and hence justifies the calculation of $\gamma$ using Eq.\,\ref{secondharmonic}.

\bigskip
\noindent
\underline{B. The field dependence}\\
\noindent
The $R^{2\omega}$ (and hence $\gamma$) measured at the EuO/KTO interface exhibits complicated evolution upon varying $H$: As displayed in main text Figs.\,3d and e, $R^{2\omega}$ changes its sign at characteristic fields $
H_1^*$ and $H_2^*$; as a result, the low-$H$ negative $R^{2\omega}$ (which is highly nonmonotonic in $H$ with a downward maximum) is replaced by a positive signal in the intermediate range of $H$, before the occurrence of a second sign reversal to negative taking place at $H_2^*$ (main text Figs.\,4a and b). This same trend was observed for both the [11$\bar{2}$] and the [1$\bar{1}$0] channels, with differences only in the characteristic temperatures and fields. We noticed that similar field-induced sign reversal of current rectification has been reported in several 2D superconducting systems \cite{ItahashiSTOSI,MasukoSI,FaxianXiu,Bi2Te3-FeTeSI} (though mostly with only one sign change). In these works, the positive and negative parts of $R^{2\omega}$ are frequently assigned to the contributions of two types of vortex. For instance, the nonreciprocal vortex motions in the activated pinned vortex states and vortex liquid states in NbSe$_2$ thin flakes prompt negative and positive $R^{2\omega}$, respectively \cite{FaxianXiu}; both vortex phases of matter are generated by an out-of-plane $H$ but they prevail in different regimes in the ($H,T$) phase diagram. Alternatively, a Bi$_2$Te$_3$/PdTe$_2$ heterostructure develops sign-changing $R^{2\omega}$ under a tilted $H$ \cite{MasukoSI} that plays a dual role on the nonreciprocity: the in-plane component triggers the current-induced renormalization of superfluid density, giving rise to a positive $R^{2\omega}$ stemming from the flow of thermally-excited free vortices, whereas the out-of-plane component induces additional vortices which are prone to the ratchet-type potential during their motions, causing a negative $R^{2\omega}$ at higher $H$. Here we note that the above-mentioned interpretations are inapplicable to this study due to the following reasons:

(i) We performed measurements with $H$ applied in-plane and the field-induced vortices are essentially absent, hence we do not expect multiple dissipative states with distinct vortex dynamics.

(ii) Even on a surface with threefold rotational symmetry, the ratchet effect engendered by disorder potential is unable to influence the flow of thermally created vortices, because the equal number of vortices and antivortices guarantees a cancellation of its impact.

(iii) In practice, a small sample misalignment may yield a finite out-of-plane component of $H$, and the induced vortices can give rise to a $R^{2\omega}$ with opposite sign. However, in this tilted-$H$ scenario, the sign change of $R^{2\omega}$ happens on a contour line $R^{\omega}/\mid H_{\perp} \mid$ = const. \cite{MasukoSI} (because samples were fixed during experiments, it also means $R^{\omega}/\mid H_{\perp} \mid$ = const.). Such prediction is in sharp contrast with our results that the sign change occurs at where $R^{\omega} \approx$ const., as demonstrated by the contour maps of both $R^{2\omega}$ (main text Figs.\,4a and b) and $\gamma$ (Supplementary Figs.\,10a and b).

Therefore, the sign-reversal features of nonreciprocal charge transport at the EuO/KTO interface are not resulted from multiple types of vortex motion. Instead, we propose that the nonmonotonic $H$ dependence of the current-induced variation in the superfluid density is responsible for the sign change at low $H$, whilst the one occurring at higher $H$ regime arises from competition between distinct mechanisms governing $R^{2\omega}$ on either side of the mean-field transition point.

The Ginzburg-Landau free energy of a 2D Rashba superconductor under an in-plane Zeeman field \textbf{H} can be written as:
\begin{equation}
F=\{ \alpha_0(H^2)+\alpha_1(H)(\hat{\bm h} \times \bm p)\cdot \hat{\bm z}+[\alpha_2(H^2)+\alpha_3(H)(\hat{\bm h} \times \bm p)\cdot \hat{\bm z}] \left| \bm p \right|^2\} \left| \Psi_{\bm{p}} \right|^2+\frac{1}{2}\beta \left| \Psi_{\bm{p}} \right|^4,
\label{S8}
\end{equation}
where $\Psi_{\bm{p}}$ is the superconductivity order parameter, $p$ is the momentum of Cooper pairs ($p \neq$ 0 is promised by the Rashba splitting of Fermi surfaces) and $\bm{\hat h} \equiv$ \textbf{H}/$H$.
The time-reversal operation dictates $\alpha_{0,2}(-H)=\alpha_{0,2}(H)$ and $\alpha_{1,3}(-H)=-\alpha_{1,3}(H)$. The influence of lattice anisotropy is not important here and thus neglected. When $\alpha_3 \neq0$, Eq.\,\ref{S8} leads to nonreciprocal electrical conductance through two different mechanisms: One is the asymmetric flow of the vortices/antivortices above $T_{\rm BKT}$ (and below $T_{\rm c}^{\rm MF}$); the other is the (Gaussian-type) amplitude fluctuations of $\Psi_{\bm{p}}$ near the mean-field transition temperature $T_{\rm c}^{\rm MF}$ \cite{HoshinoSI,ItahashiSTOSI}.
To understand the nonreciprocal vortex flow, we follow the approach in Ref.\,\cite{HoshinoSI} and replace the momentum $\bm p$ after $\alpha_3$ by the average value $<\bm p>= p_s\bm{\hat x} \sim \bm I$. Also, the $\alpha_1$ term may be neglected for now, since it can be eliminated by shifting the zero point of $\bm p$. Then Eq.\,\ref{S8} becomes:
\begin{equation}
F=\{ \alpha_0(H^2_y)+[\alpha_2(H^2_y)+\alpha_3(H_y)p_s]\bm p^2 \} \left| \psi \right|^2+\frac{1}{2}\beta \left| \psi \right|^4.
\label{S9}
\end{equation}
Thus, the superfluid density,
\begin{equation}
n_s \sim \alpha_2(H^2_y)+\alpha_3(H_y)p_s,
\label{n_s}
\end{equation}
which obviously varies with the current $I$, and thus $\bm{p}_s$, is reversed. Since $T_{\rm BKT} \sim n_s$, Eq.\,\ref{n_s} suggests that the BKT transition temperature shows nonreciprocity, leading to nonreciprocal resistance through the relation $R_{\rm s} \sim {\rm exp}\{ -b\sqrt{(T_{\rm c}^{\rm MF}-T_{\rm BKT})/(T-T_{\rm BKT})}\}$. In general, $\alpha_3(H_y)$ is a nonmonotonic function of $H_y$, instead of being $H$-linear in the treatment of Ref.\,\cite{HoshinoSI}. Our numerical calculations (Supplementary Fig.\,10c) unequivocally shows that $\alpha_3(H_y)$ can change its sign as $H_y$ increases, rendering sign reversal of the nonreciprocal $R^{2\omega}$ signal. Furthermore, the sign reversal occurs at a field $H^{\ast}_1$ decreasing as the temperature is lowered. Both the above features are consistent with the observation of $R^{2\omega}(H)$ at small fields in main text Figs.\,3d and e.

A second sign reversal of $R^{2\omega}$ happens at a higher characteristic field $H_2^{\ast} \sim \sqrt{T_{\rm c}^{\rm MF}(0)-T}$; such behavior traces the evolution of mean-field in-plane upper critical field $H_{\rm c2}$ (Eq.\,\ref{Hc2para}), suggesting that $H_2^{\ast}$ is delineated by the suppressed mean-field transition temperature under the Zeeman field $H$. Indeed, we define $H_{\rm c2}$ using the 50$\%$\,$R_{\rm N}$ criterion (Supplementary Fig.\,5) and $H_2^{\ast}$ is close to the contour line of $R = R_{\rm N}/3$ (main text Fig.\,4), thus $H_2^{\ast} \sim (2/3)H_{\rm c2}$ is roughly satisfied. In a 2D superconductor with low superfluid stiffness such as the KTO-based interfaces \cite{KTOsuperfluidSI}, the in-plane $H_{\rm c2}$ is more likely a continuous crossover rather than a distinct phase transition \cite{VortexReview}. In the vicinity of this crossover between the superconducting and normal states, there exist a region with a finite width wherein the fluctuation in the amplitude of $\Psi_{\bm{p}}$ (i.e., the same Gaussian-type fluctuations referred to in the AL model) play a substantial role determining the physical properties \cite{VortexReview,GalitskiLarkin}. As such, we postulate that the sign change of $R^{2\omega}$ at $H_2^{\ast}$ reflects that the vortex-flow contribution at lower $H$ is overcame by the contribution of amplitude fluctuations; this explanation is also supported by the fact that $R^{2\omega}$ at $H > H_2^{\ast}$ has the same sign (minus) as $T > T_{\rm c}^{\rm MF}(0)$ (main text Figs.\,4a and b). Strong amplitude fluctuation effects in magnetoresistance below the nominal $H_{\rm c2}$ has also been observed in cuprate superconductors \cite{YBCO-MR1,YBCO-MR2}.

On a final note, we emphasize that only a preliminary picture is presented here to aid in understanding the observed nonreciprocal properties. Further investigation is necessary to fully elucidate the underlying mechanisms.

\bigskip

\noindent
\textbf{Note 7: One-dimensionality of the putative superconducting stripes}
\par
\noindent
Based on the data collected on EuO/KTO(110) interfaces, we have proposed a ``superconducting stripe" scenario in which a self-organized phase separation develops at the 2D interface, the superconducting phase with higher $T_{\rm c}$ forms quasi-one-dimensional (quasi-1D) stripe-like structures running across the interface along the [001] axis; considering that the scanning SQUID method did not directly resolve these stripes, we conjecture that their widths are at a sub-micrometer level \cite{HuaStripeSI}. Akin to the EuO/KTO(110) interfaces, here we reveal that anisotropic superconducting transitions occur also at the EuO/KTO(111) interface; if this originates from similar superconducting stripes, these channels must extend along one specific direction among the three equivalent [11$\bar{2}$] axes. An immediate question is whether such stripes behave like 1D superconductors. In 1D systems, long-range phase coherence of superconductivity is absent; the superconducting order parameter is prone to thermal and quantum fluctuations, with its amplitude having a finite probability to become zero at at certain positions \cite{Little,Rogachev}. Along a 1D superconducting wire, these zero points of superconducting order parameter correspond to a series of elementary excitations \cite{CaZrO3StonerSI} at which the phase of the order parameter to slip by 2$\pi$, i.e., the phase slips \cite{Rogachev,CNLau}. The thermally-activated phase slips causes emergent voltage signals, giving rise to a unique form of the resistance for a superconducting wire that is described by the Langer-Ambegaokar-McCumber-Halperin (LAMH) theory \cite{LangerAmbegaokar,McCumberHalperin}:
\begin{equation}
R_{\rm LAMH}(T)= \frac{\pi\hbar^2\Omega}{2e^2k_{\rm B}T}\exp{[-\frac{\Delta F(T)}{k_{\rm B}T}]},
 \label{LAMH}
\end{equation}
\begin{equation}
\Delta F(T)=\frac{8\sqrt{2}}{3}\frac{H^2_{\rm c}(T)}{8\pi}S\xi_{\rm GL}(T),\\
\end{equation}
\begin{equation}
\Omega=\frac{L}{\xi_{\rm GL}(T)\tau_{\rm GL}}[\frac{\Delta F(T)}{k_{\rm B}T}]^{-1/2},\\
\end{equation}
\begin{equation}
\tau_{\rm GL}=\pi\hbar/8k_{\rm B}(T_{\rm c}-T)
\end{equation}
\noindent
here $\Delta F(T)$ is the energy barrier between two neighboring metastable states to be overtaken by thermal fluctuations, $\Omega$ is the attempt frequency, $\tau_{\rm GL}$ is the characteristic relaxation time in the time dependent GL theory, $H_{\rm c}(T)=H_{\rm c}(0)(1-T/T_{\rm c})$ is the thermodynamic critical field, $\xi_{\rm GL}(T)=\xi_{\rm GL}(0)(1-T/T_{\rm c})^{-1/2}$ is the GL coherence length. $L$ and $S$ are the length and cross-sectional area of the wire, respectively, and $k_{\rm B}$ the Boltzmann constant \cite{Tinkham,Rogachev,CNLau}. In addition, as the parameter $H_{\rm c}$ is ambiguous in the present study, we adopted an alternative expression of the (approximation of) energy barrier \cite{CNLau}:
\begin{equation}
\Delta F(T) \approx 0.83k_B T_{\rm c}\frac{L}{\xi_{\rm GL}(0)}\frac{R_{\rm q}}{R_{\rm N}}(1-\frac{T}{T_{\rm c}})^{3/2},
 \label{barrier}
\end{equation}
in which $R_{\rm q} = h/4e^2$ = 6.45\,k$\Omega$ is the resistance quantum for superconducting states. In practice, the measured resistance of the superconducting wire is fitted to $R(T)$ = $[(R_{\rm N})^{-1}+(R_{\rm LAMH})^{-1}]^{-1}$.

Application of the LAMH model (Eqs.\,\ref{LAMH}-\ref{barrier}) fails to reproduce the experimental data of $R(T)$ across the superconducting transition measured along the $[11\bar{2}]$ channel with the highest $T_{\rm c}$ in Device 2 (cyan color in Supplementary Fig.\,8a), as illustrated in Supplementary Fig.\,11. Using the experimentally determined parameters of $R_{\rm N}$, $\xi_{\rm GL}(0)$ and $T_{\rm c}^{\rm MF}$, $\Delta F$ at $T$ = 0\,K is calculated to be $\Delta F(0)$ = 2.19$\times$10$^{-18}$\,J for this channel. Nevertheless, if we select the value of $L$ to be the length of the Hall-bar channel (600\,$\mu$m), the LAMH model yields a resistance correction that profoundly deviates from the data (dotted lines in Supplementary Fig.\,11). On the other hand, once we set $L$ = 160\,nm (a typical length for superconducting nanowires), the LAMH fits suffice to trace the data between 1.2 and 1.7\,K (solid line in Supplementary Fig.\,11). We conclude that if the superconducting stripes do run across the full length of the channel (as assumed in our original picture \cite{HuaStripeSI}), they can hardly be regarded as superconducting nanowires; the length scale of several hundred micrometers is too long for the establishment of 1D superconductivity. Hence, we postulate that these stripes, if exist, may still have a 2D nature with their width exceeding the superconducting coherence length ($\simeq$ 20\,nm). In this scenario, what occurs at the interface is an exotic phase segregation giving rise to two 2D superconducting regimes, the one with higher $T_{\rm c}$ form a fully connected path only along the specific $[11\bar{2}]$ direction.

\bigskip

\textbf{SUPPLEMENTARY REFERENCES}

\newpage

\setcounter{figure}{0}
\counterwithin*{figure}{part}
\renewcommand{\thefigure}{S\arabic{figure}}

\begin{figure*}[htbp!]
	\centering
	\includegraphics[width=1\textwidth]{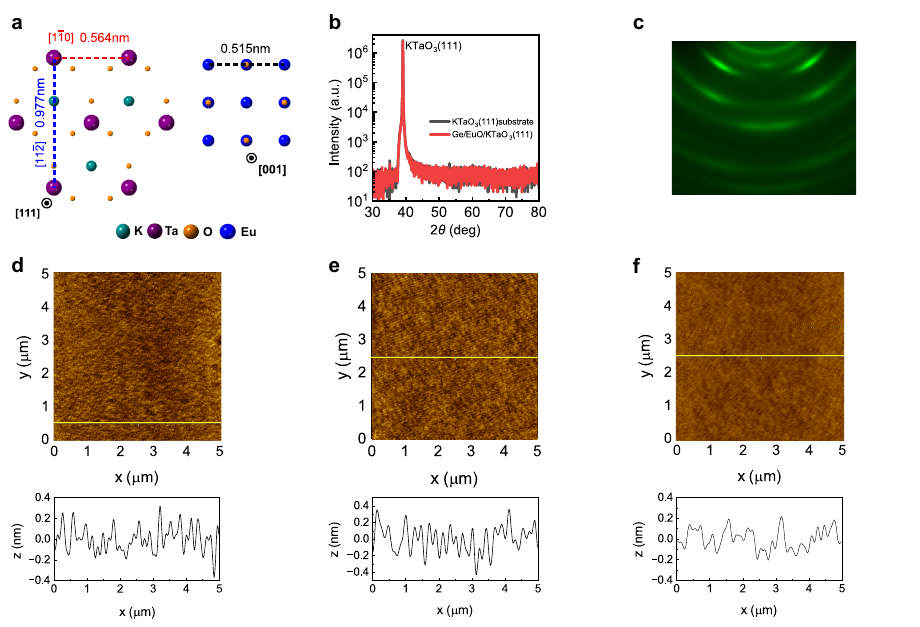}
	\caption{\textbf{Structural characterization of the EuO/KTaO$_3$(111) interfaces.} \textbf{a}, Illustrations of the KTO(111) plane and the EuO(001) plane showing the positions of different ions. The distance between nearest Ta ions along $[1\bar{1}0]$ ($[11\bar{2}]$) is 0.564 (0.977 nm). The distance between nearest Eu ions is 0.515 nm. \textbf{b}, $\theta-2\theta$ X-ray diffraction (XRD) patterns of a (111)-oriented KTO substrate (black) and a Ge/EuO/KTO heterostructure (red). The EuO film does not contribute any visible Bragg peak. \textbf{c}, The ring-shaped reflective high-energy electrons diffraction (RHEED) pattern of the EuO epitaxial film demonstrates the polycrystalline nature of the overlayer. \textbf{d-f}, Atomic force microscope (AFM) images measured on two samples with $n_{\rm 2D} = 9.79\times10^{13}$ cm$^{-2}$ (\textbf{d}) and 7.67$\times$10$^{13}$ cm$^{-2}$ (\textbf{e}), respectively, as well as an annealed KTO substrate (\textbf{f}). Scanned areas have the size of 5$\times$5\,$\mu$m$^{2}$. The surface roughness (see curves in the bottom panels) was measured along the horizontal yellow bar in the AFM images; the typical root mean square roughness $R_{\rm q}$ is 3.5, 4.4 and 2.7\,\AA\, for (\textbf{d}), (\textbf{e}) and (\textbf{f}), respectively.
	}
	\label{Fig.s1}
\end{figure*}

\begin{figure*}[htbp!]
	\centering
	\includegraphics[width=1\textwidth]{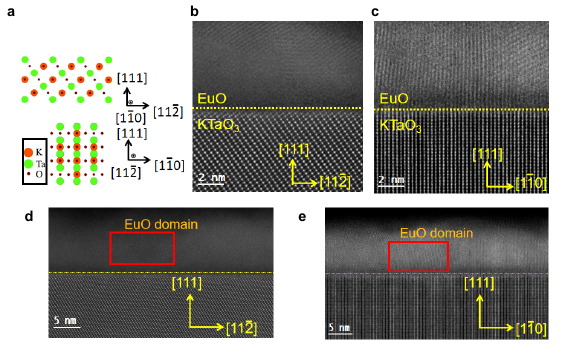}
	\caption{\textbf{STEM images of EuO/KTO(111) interfaces.} \textbf{a}, Schematics showing the atomic arrangement of KTO in the (1$\bar{1}$0) and (11$\bar{2}$) planes. \textbf{b}, \textbf{c}, Cross-sectional atomic-resolution STEM images of a EuO/KTO(111) interface (Sample No.\,2) looking down the KTO [1$\bar{1}$0] (\textbf{b}) and [11$\bar{2}$] (\textbf{c}) crystallographic directions, respectively. Yellow dashed lines mark the position of interface. \textbf{d}, \textbf{e}, Low-magnification STEM images viewed along the KTO [1$\bar{1}$0] (\textbf{d}) and [11$\bar{2}$] (\textbf{e}) directions, respectively. Red box roughly corresponds to a polycrystal grain of the EuO film.
	}
	\label{Figs2}
\end{figure*}
\clearpage

\begin{figure*}[htbp!]
	\centering
	\includegraphics[width=0.75\textwidth]{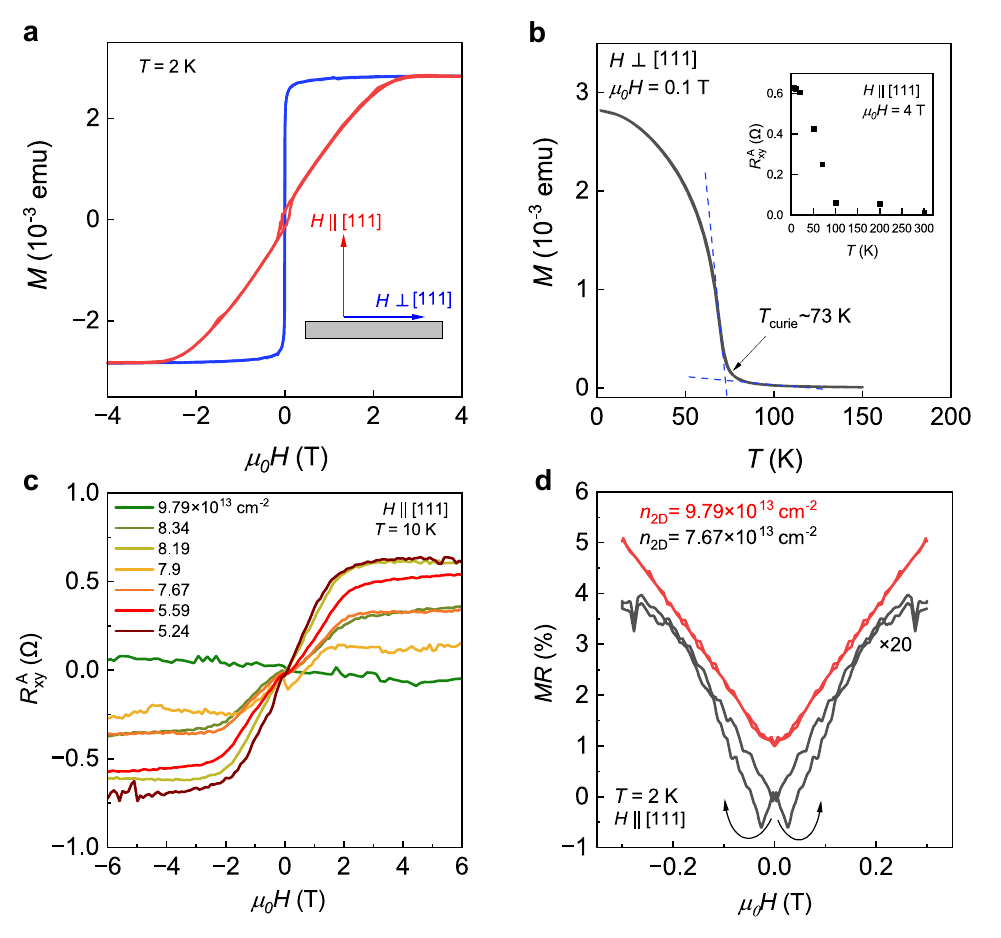}
	\caption{\textbf{Magnetic properties of the EuO/KTaO$_3$(111) interfaces.} \textbf{a}, Magnetization $M(H)$ measured at $T$ = 2\,K under in-plane (blue) and out-of-plane (red) magnetic fields. The signal is predominantly contributed by the ferromagnetic EuO overlayer. \textbf{b}, $T$-dependent magnetization $M$ measured during a zero-field-cool sweep under an in-plane $\mu_0H$ = 0.1\,T. The Curie temperature is determined to be $T_{\rm Curie} \approx$ 73\,K. Inset shows the $T$ dependence of the (saturated) anomalous Hall resistance $R_{xy}^{A}$, whose onset coincides with $T_{\rm Curie}$. \textbf{c}, $R_{xy}^{A}(H)$ measured at $T =$ 10\,K for interfacial 2DEGs with different $n_{\rm 2D}$ listed in Supplementary Table 1; $R_{xy}^{A}$ was obtained by subtracting the $H$-linear ordinary term from $R_{xy}$. \textbf{d}, Magnetoresistance (MR, $[R(H)-R(H=0)]/R(H=0)\times100\%$) measured at 2 K under in-plane $H$. Red and black curves are measured on the interface samples No. 7 and No. 1, with $n_{\rm 2D}$ = 9.79 and 7.67$\times$10$^{13}$ cm$^{-2}$, respectively.
	}
	\label{Figs3}
\end{figure*}

\begin{figure*}[htbp!]
	\centering
	\includegraphics[width=1\textwidth]{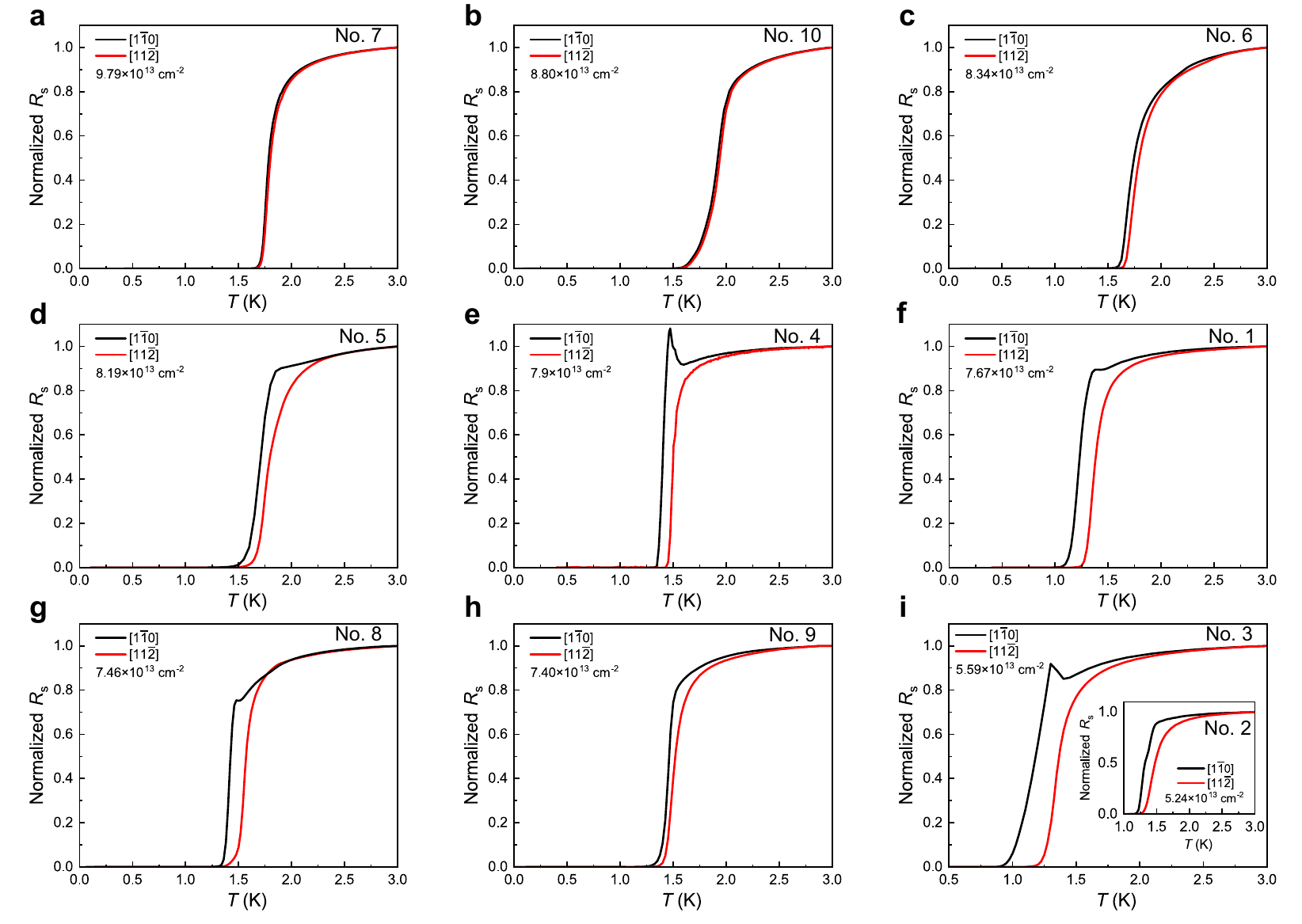}
	\caption{\textbf{Van der Pauw measurements on samples with different $n_{\rm 2D}$.} The normalized sheet resistance $R_{\rm s}(T)/R_{\rm s}$(3 K) as a function of $T$, measured in the van der Pauw configuration [inset of (\textbf{a})] on EuO/KTO(111) 2DEGs with different $n_{\rm 2D}$: \textbf{a}, $n_{\rm 2D}$ = 9.79$\times$10$^{13}$ cm$^{-2}$ (Sample No.\,7). \textbf{b}, $n_{\rm 2D}$ = 8.80$\times$10$^{13}$ cm$^{-2}$ (Sample No.\,10). \textbf{c}, $n_{\rm 2D}$ = 8.34$\times$10$^{13}$ cm$^{-2}$ (Sample No.\,6). \textbf{d}, $n_{\rm 2D}$ = 8.19$\times$10$^{13}$ cm$^{-2}$ (Sample No.\,5). \textbf{e}, $n_{\rm 2D}$ = 7.90$\times$10$^{13}$ cm$^{-2}$ (Sample No.\,4). \textbf{f}, $n_{\rm 2D}$ = 7.67$\times$10$^{13}$ cm$^{-2}$ (Sample No.\,1). \textbf{g}, $n_{\rm 2D}$ = 7.46$\times$10$^{13}$ cm$^{-2}$ (Sample No.\,8).\textbf{h}, $n_{\rm 2D}$ = 7.40$\times$10$^{13}$ cm$^{-2}$ (Sample No.\,9). \textbf{i}, $n_{\rm 2D}$ = 5.59$\times$10$^{13}$ cm$^{-2}$ (Sample No.\,3). For the black (red) curves, the current $I$ is applied along the crystallographic [1$\bar{1}$0] ([11$\bar{2}$]) direction. Inset shows data of the sample with $n_{\rm 2D}$ = 5.24$\times$10$^{13}$ cm$^{-2}$ (on which Device 2 was fabricated).
	}
	\label{Figs4}
\end{figure*}
\clearpage

\begin{figure*}[htbp!]
	\centering
	\includegraphics[width=1\textwidth]{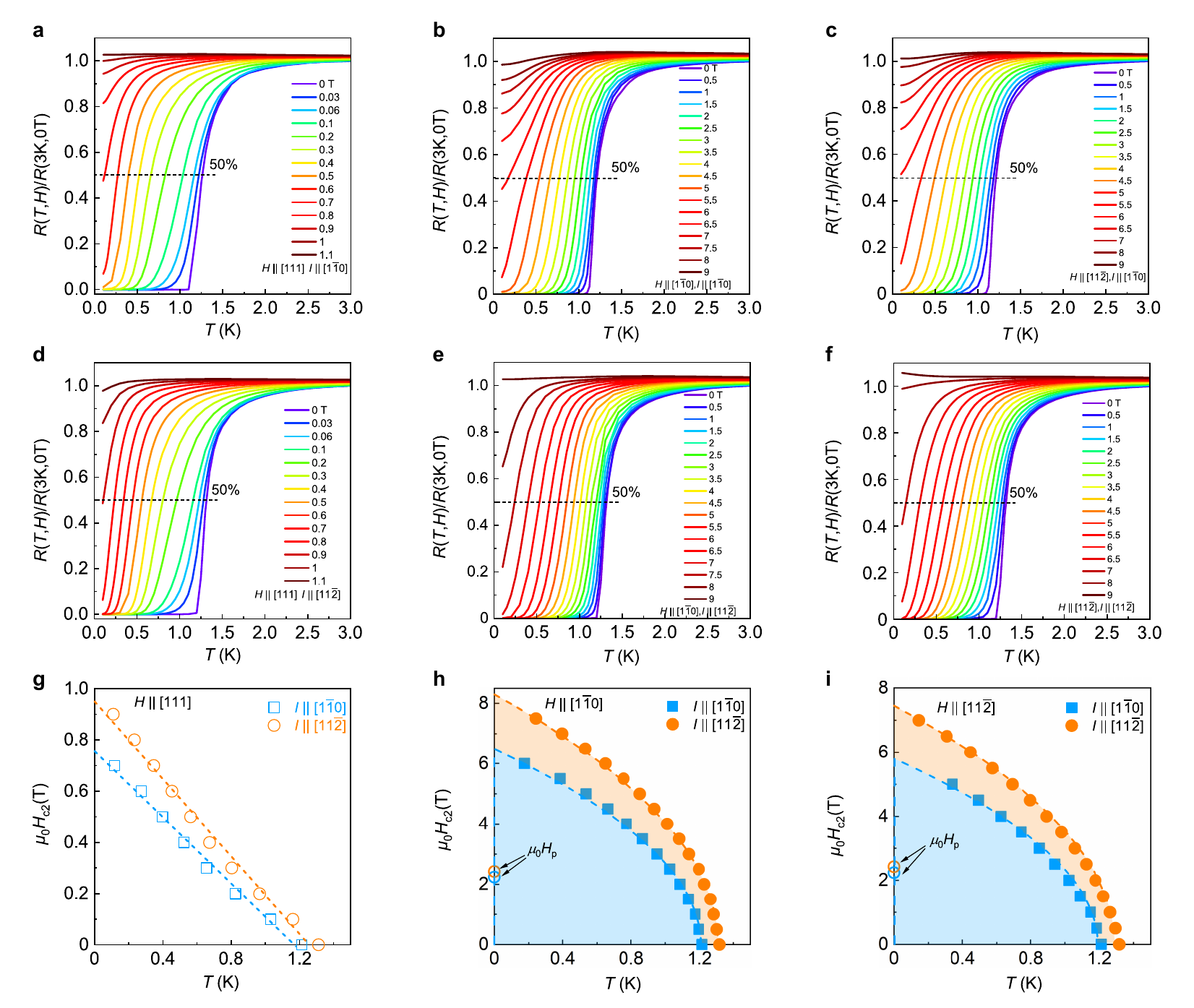}
	\caption{\textbf{Upper critical fields.} \textbf{a-c}, Normalized sheet resistance plotted against $T$ measured on Device 1 ($n_{\rm 2D} = 7.67\times10^{13}$\,cm$^{-2}$) with $I \parallel$ [1$\bar{1}$0] and $H \parallel$ [111] (\textbf{a}), $H \parallel$ [1$\bar{1}$0] (\textbf{b}) and $H \parallel$ [11$\bar{2}$] (\textbf{c}). Different colors denote varying $H$. \textbf{d-f}, Same as \textbf{a-c} but with $I \parallel$ [11$\bar{2}$]. Dotted horizontal lines in (\textbf{a})-(\textbf{f}) marks the position where the resistance drops to 50$\%$ of its normal state value (at $T$ = 3\,K). \textbf{g-i}, The upper critical field $H_{\rm c2}$ as a function of $T$ determined for Device 1: (\textbf{g}), $H \parallel$ [111]; (\textbf{h}), $H \parallel$ [1$\bar{1}$0]; (\textbf{i}), $H \parallel$ [11$\bar{2}$]. $H_{\rm c2}$ values were determined from the data shown in (\textbf{a})-(\textbf{f}) using the 50$\%$ criterion (dotted lines). Orange circles and blue squares represent $H_{\rm c2}$ for the current direction along [11$\bar{2}$] and [1$\bar{1}$0], respectively. Dashed lines are the fits based on the Ginzburg-Landau theory for a 2D superconductor \cite{Tinkham}. The Pauli paramagnetic limit $\mu_0H_{\rm p} \approx 1.76k_{\rm B}T_{\rm c}$ is marked on the vertical axis for both current directions.
	}
	\label{Figs5}
\end{figure*}

\begin{figure*}[htbp!]
	\centering
	\includegraphics[width=1\textwidth]{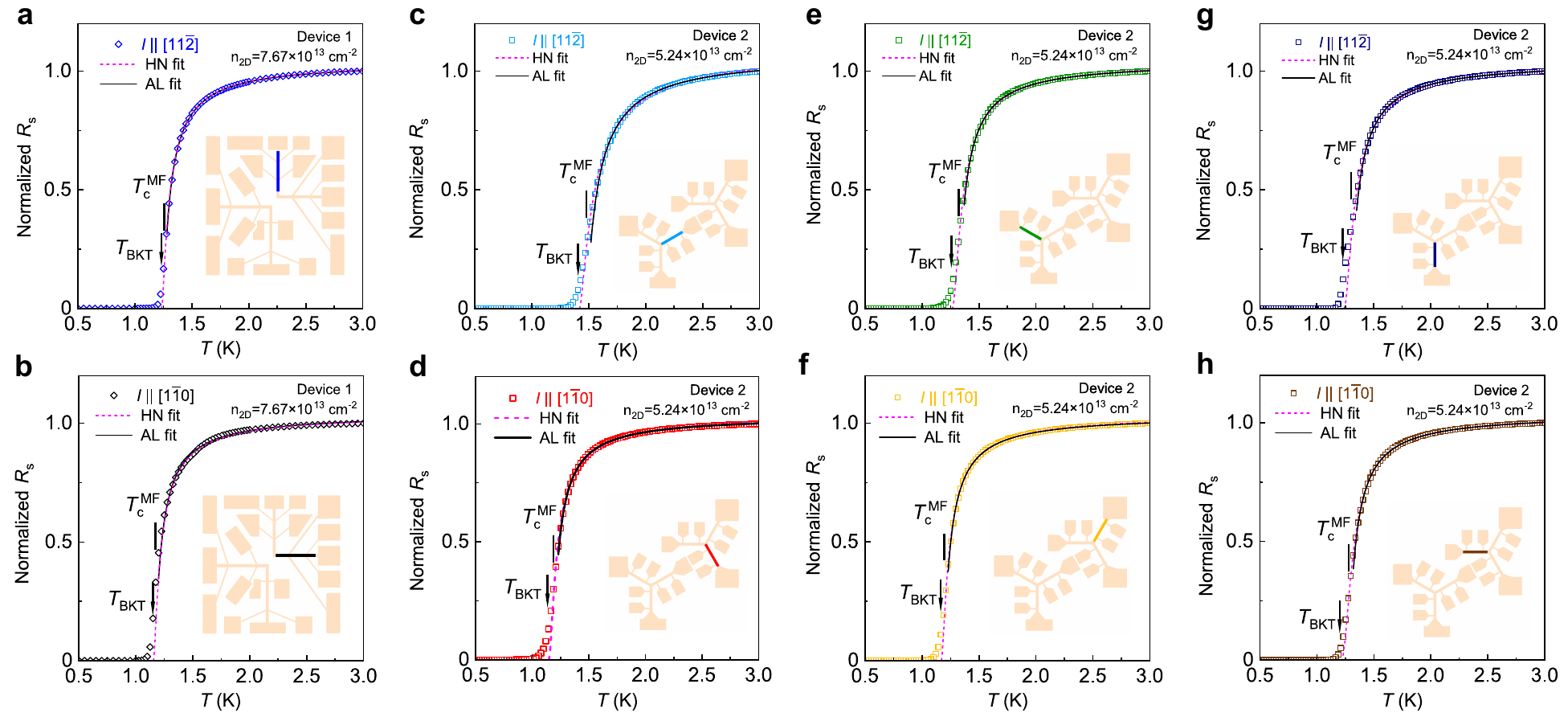}
	\caption{\textbf{Analysis of the paraconductivity.} The normalized $R_{\rm s}$ (open symbols) and the corresponding fitting curves using the Halperin-Nelson (HN) formula (Eq.\,\ref{HN}, dashed lines) and the Aslamazov-Larkin (AL) model (Eq.\,\ref{ALfit}, solid lines). The fitting results of $T_{\rm BKT}$ and $T_{\rm c}^{\rm MF}$ are indicated by arrows and vertical bars, respectively. The AL model fits the data curves well for $T > T_{\rm c}^{\rm MF}$. A resistive ``tail" appearing below $T_{\rm BKT}$ marks the deviation from the HN model, which originates from inhomogeneity and/or finite-size effect (see text). Data are presented for the [11$\bar{2}$] channels (\textbf{a,c,e,g}) and the [1$\bar{1}$0] channels (\textbf{b,d,f,h}) in Device 1 (\textbf{a} and \textbf{b}, blue and black diamonds) and Device 2 (\textbf{c-h}, colors of symbols denote the Hall-bar channel on which the data were taken; the correspondence is the same as in main text Fig.\,2 and Supplementary Fig.\,8a).
	}
	\label{Figs6}
\end{figure*}

\begin{figure*}[htbp!]
	\centering
	\includegraphics[width=1\textwidth]{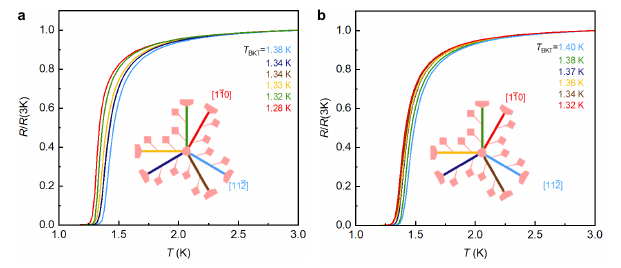}
	\caption{\textbf{Data taken from the radial Hall-bar devices.} Normalized $R_{\rm s}$ curves measured on Device 4 (\textbf{a}) and Device 5 (\textbf{b}), fabricated on Samples No.\,8 and No.\,9, respectively. These two devices share the same design composed of six Hall bars (along three $[11\bar{2}]$ and three $[1\bar{1}0]$ directions, respectively) with one end of each concentrating on the same spot located near the center of the sample, see the sketches in the insets. Colours of the curves and the Hall-bar channels are set in correspondence. The BKT transition temperature for each channel is listed in Supplementary Table 4.
	}
	\label{Figs7}
\end{figure*}

\begin{table*}[htb]
	\renewcommand{\thetable}{S4}
	\renewcommand*{\thetable}{S\arabic{table}}
	\centering
	\caption{\textbf{BKT transition temperature measured on Devices 4 and 5.} $T_{\rm BKT}$ was obtained by fitting the Halperin–Nelson formula to the resistance curves shown in Supplementary Fig.\,7. Note that aside from the $[11\bar{2}]$ channel with the highest $T_{\rm BKT}$ (cyan) and the $[1\bar{1}0]$ channel perpendicular to it showing the lowest $T_{\rm BKT}$ (red), there is no explicit ordinal relationship among $T_{\rm BKT}$ measured along the remaining four channels.}
	\label{tab4}
	\begin{ruledtabular}
		\begin{tabular}{c|c|c|c|c|c|c}
			Device No.  & \multicolumn{3}{c|}{$[11\bar{2}]$ channels}  & \multicolumn{3}{c}{$[1\bar{1}0]$ channels} \\\hline
			& cyan & green & navy & red & amber gold & brown \\\hline
			4 & 1.38 K & 1.32 K & 1.34 K & 1.28 K & 1.33 K & 1.34 K \\\hline
			5 & 1.40 K & 1.38 K & 1.37 K & 1.32 K & 1.34 K & 1.36 K \\
		\end{tabular}
	\end{ruledtabular}
\end{table*}

\begin{figure*}[htbp!]
	\centering
	\includegraphics[width=1\textwidth]{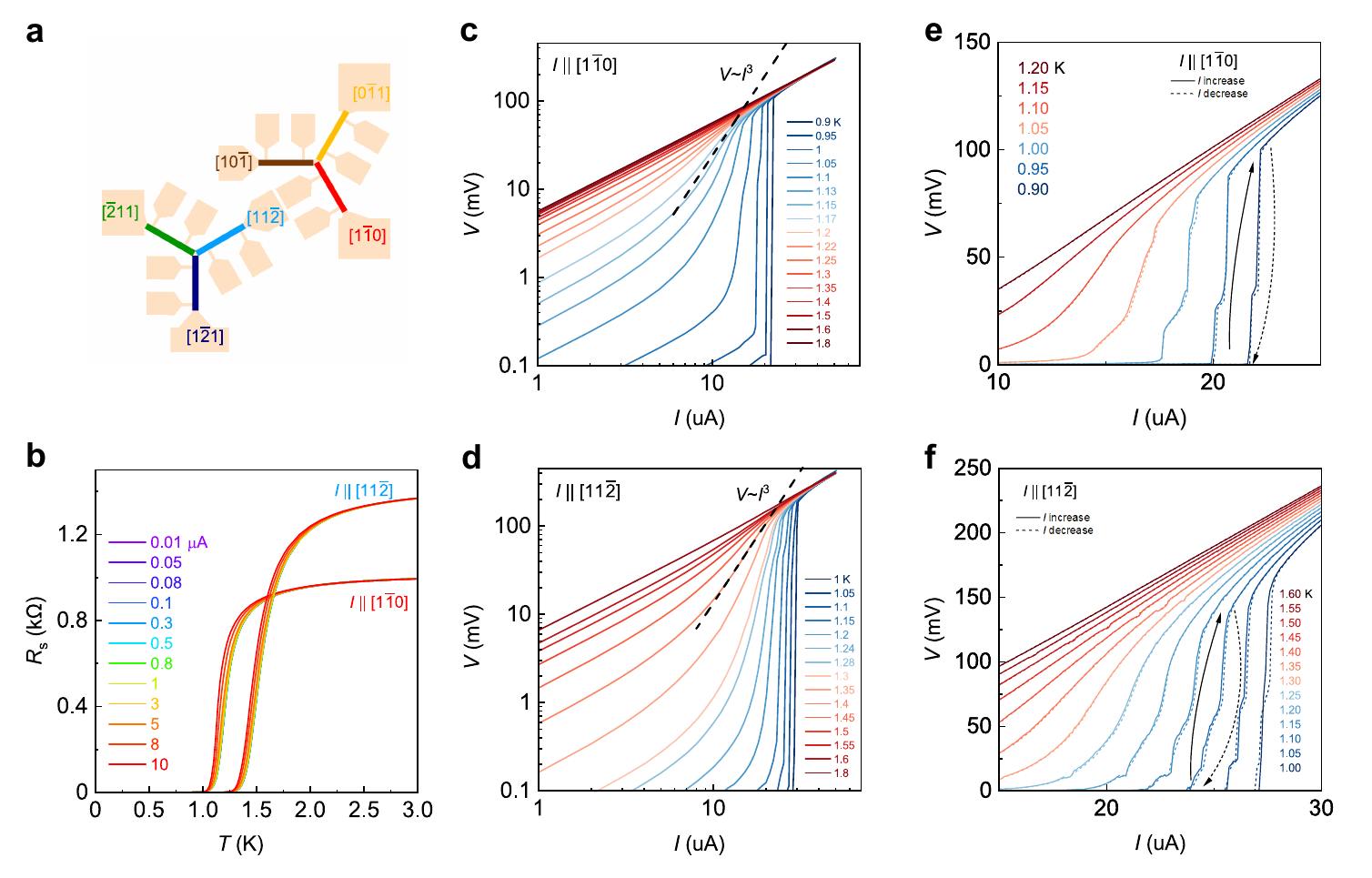}
	\caption{\textbf{The $I-V$ characteristics.} \textbf{a}, Schematics showing the double-tri-beam Hall-bar pattern for Device 2. This device was fabricated on one EuO/KTO interface, enabling the transport measurements along six different directions on the same sample. The corresponding crystal axes are labeled on the Hall bars. \textbf{b}, $R_{\rm s}(T)$ measured on one [11$\bar{2}$] channel (cyan) and one [1$\bar{1}$0] channel (red) in Device 2 with current $I$ varying from 0.01 $\mu$A to 10 $\mu$A. The superconducting transitions show insignificant current dependence up to approximately 1 $\mu$A, thus the influence of current-induced excitation can be neglected for lower $I$. \textbf{c},\textbf{d}, $T$-dependent $I-V$ curves measures along \textbf{c}, the [1$\bar{1}$0] channel [red in \textbf{a}] and (\textbf{d}) the [11$\bar{2}$] [blue in \textbf{a}]; data are shown for the sweeping range of $I$ from 1 to 50 $\mu$A. Dashed lines indicate the $V \propto I^3$ relationship expected at $T_{\rm BKT}$. \textbf{e},\textbf{f}, Expanded view of the data presented in (\textbf{c}) and (\textbf{d}), highlighting the clockwise hysteresis between up and down sweeps of $I$ (solid and short-dashed curves, respectively).
    }
	\label{Figs8}
\end{figure*}

\begin{figure*}[htbp!]
	\centering
	\includegraphics[width=1\textwidth]{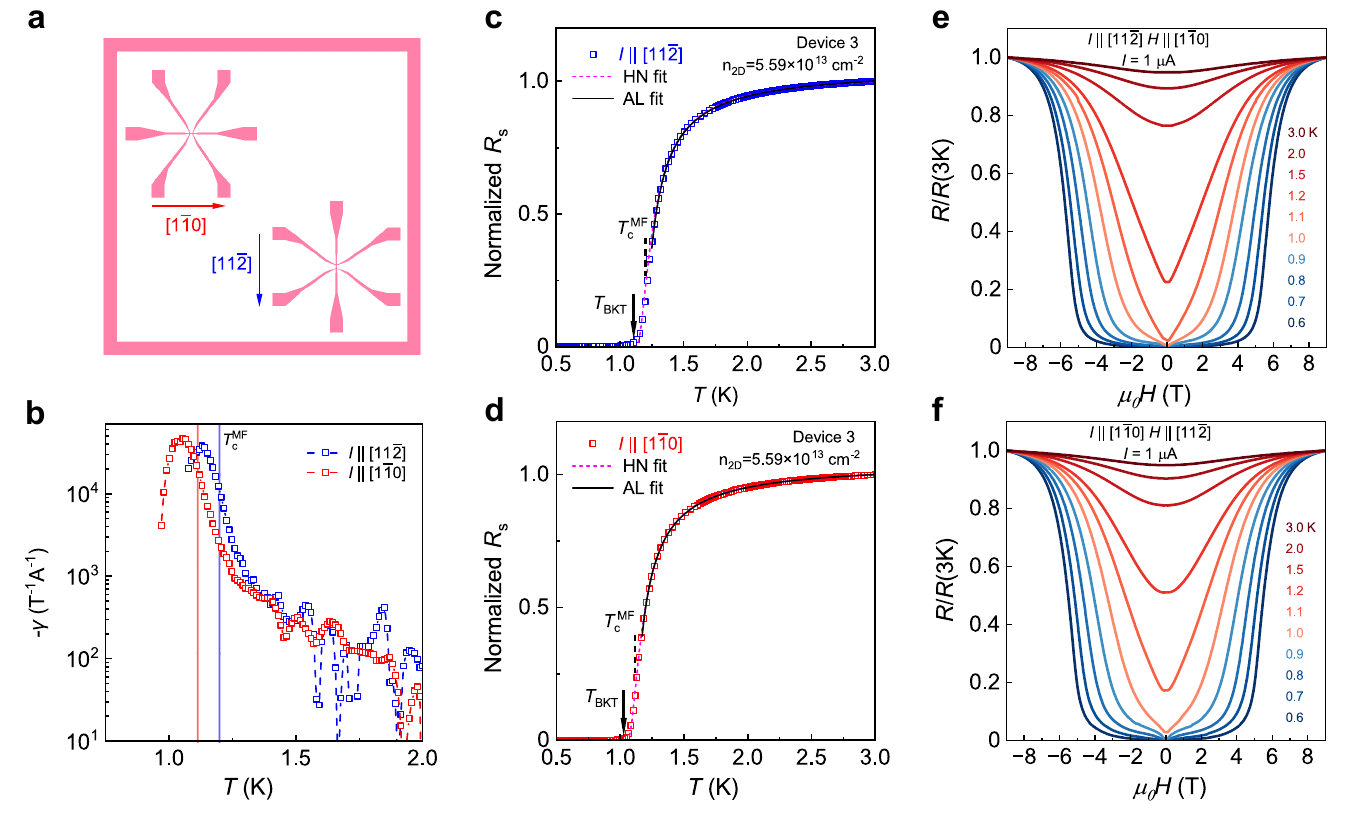}
	\caption{\textbf{Transport measurements on Device 3.} \textbf{a}, Schematic drawing of the configuration for Device 3: Two Hall-bar channels were fabricated along the [11$\bar{2}$] and [1$\bar{1}$0] directions (arrows), respectively. \textbf{b}, The coefficient $\gamma$ for current rectification plotted in a logarithmic scale against $T$. Blue (red) symbols denote data collected on the [11$\bar{2}$] ([1$\bar{1}$0]) channel; vertical lines indicate $T_{\rm c}^{\rm MF}$ for each channel. Above $\sim$ 1.6\,K, $\gamma$ is almost submerged by the experimental noise floor. \textbf{c},\textbf{d}, The Halperin-Nelson fits (dashed line) and the Aslamazov-Larkin fits (solid lines) to the normalized sheet resistance $R_{\rm s}$ (essentially the linear resistance $R^{\omega}(T)$) for the (\textbf{c}) [11$\bar{2}$] and (\textbf{d}) [1$\bar{1}$0] channels, respectively. $T_{\rm BKT}$ and $T_{\rm c}^{\rm MF}$, are marked by arrows and vertical bars, respectively. (\textbf{e,f}) $R^{\omega}(H)$ measured at different $T$ (from 0.6 to 3\,K) normalized to the value at $T$ = 3\,K and $\mu_0H$ = 9\,T. Data in (\textbf{e}) and (\textbf{f}) are shown for the [11$\bar{2}$] and [1$\bar{1}$0] channels, respectively. $H$ is applied perpendicular to $I$. All the data in this figure were measured with a bias current $I$ = 1\,$\mu$A.
	}
	\label{Figs9}
\end{figure*}

\begin{figure*}[htbp!]
	\centering
	\includegraphics[width=0.75\textwidth]{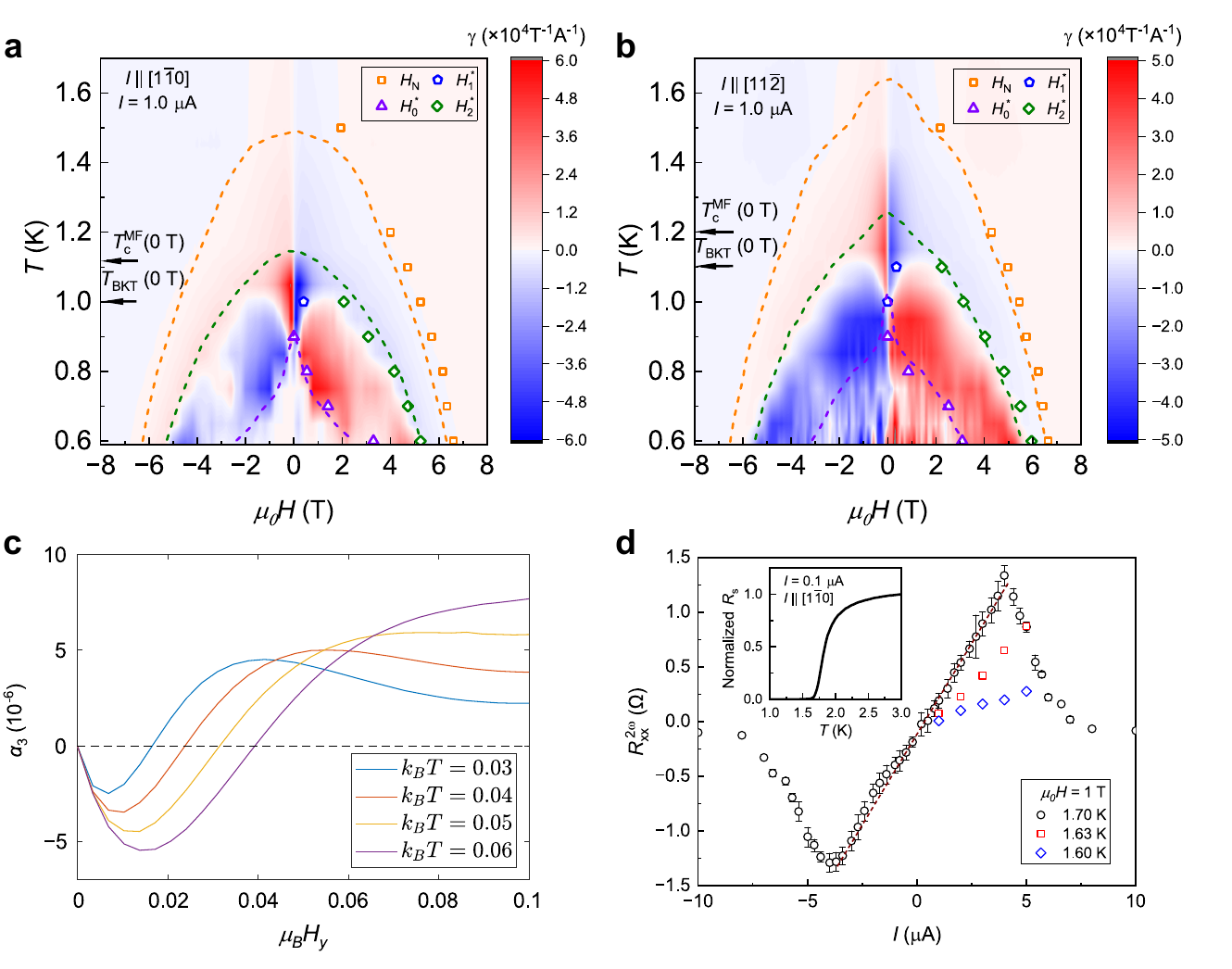}
	\caption{\textbf{Field dependence of the nonreciprocal responses in the superconducting regime.} \textbf{a},\textbf{b}, Color contour plots of $\gamma$ in the ($H,T$) plane, presented for $I \parallel$ [1$\bar{1}$0] (\textbf{a}) and $I \parallel$ [11$\bar{2}$] (\textbf{b}), respectively. Data are derived from the measurement results of $R^{2\omega}$ (main text Figs.\,3d, e) and $R^{\omega}$ (Supplementary Figs.\,9e, f): $\gamma=\frac{2R^{2\omega}(H,T)}{\mu_0HIR^{\omega}(H,T)}$. $T_{\rm BKT}$ and $T_{\rm c}^{\rm MF}$ for each direction are denoted by black arrows on the vertical axis. Symbols overlaid to the contour plots are the characteristic fields: $H_0^*$ (purple triangles), $H_1^*$ (blue pentagons), $H_2^*$ (green diamonds) and $H_{\rm N}$ (orange squares). Orange and green short-dashed lines are contour lines of $R^{\omega}$ = 0.8\,$R_{\rm N}$ and (1/3)\,$R_{\rm N}$, respectively. Purple short-dashed line denotes the contour line of $R$ = 0.01\,$R_{\rm N}$. \textbf{c}, The third-order coefficient $\alpha_3$ of the Cooper-pair kinetic term in Eq.\,\ref{S9} as a function of the Zeeman field at various temperatures. It can be seen that $\alpha_3(H)$ displays $T$-dependent minimum and maximum, as well as a sign reversal from negative to positive; these signatures qualitatively reproduce the experimentally observed profile of $R^{2\omega}(H)$ (main text Fig.\,3). Calculations are based on a normal Hamiltonian $\mathcal{H} = k^2/2m-\mu+r(k_ys_x-k_xs_y)+H_ys_y$, where $s_{1,2}$ are the Pauli matrices and $m$ = 1, $r$ = 0.5, $\mu$ = -0.02.  \textbf{d}, Current $I$ dependence of $R^{\rm 2\omega}$ measured at 1.70 K (black circles), 1.63 K (red squares) and 1.60 K (blue diamonds) under a magnetic field $\mu_0H$ = 1\,T applied along the in-plane [11$\bar{2}$] direction. Data were taken on Device 6 (Hall bar fabricated on Sample No.\,10, $n_{\rm 2D}$ = 8.8$\times$10$^{13}$ cm$^{-2}$) with $I \parallel$ [1$\bar{1}$0]. Wine dashed line indicates the linear relationship between $R^{\rm 2\omega}$ and $I$ in the low-current region. Inset: normalized $T$-dependent resistance measured on the same device.
	}
	\label{Figs10}
\end{figure*}
\clearpage

\begin{figure*}[htbp!]
	\centering
	\includegraphics[width=0.5\textwidth]{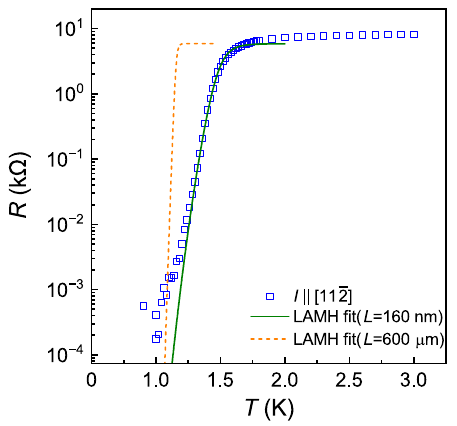}
	\caption{\textbf{Test of the 1D superconducting nanowire scenario.} The validity of the superconducting nanowire picture was examined by fitting the Langer-Ambegaokar-McCumber-Halperin (LAMH) model to the $R(T)$ data. The data (blue squares) are shown in the vicinity of the superconducting transition of the $[11\bar{2}]$ channel in Device 2 (cyan in main text Fig.\,2a). This channel possesses the highest $T_{\rm c}$, which makes more likely to harbor characteristics of 1D superconducting structures predicted by the LAMH model. Nonetheless, the LAMH fit (orange dashed line) profoundly deviates from the data if we set the length $L$ of the 1D nanowire to the full length of the channel (600\,$\mu $m). Solid green line is the LAMH fit taking a much shorter $L$ = 160 nm. It elucidates that if the channels with higher $T_{\rm c}$ contain 1D superconducting textures, the length of these textures may not exceed 100-200\,nm.
	}
	\label{Figs11}
\end{figure*}


\noindent
\textbf{References}

\begin{thebibliography}{99}
	
	
	
	


\bibitem{Pearl}
Pearl, J. in {\it Low Temperature Physics-LT9}, edited
by Daunt, J. G., Edwards, D. O., Milford, F. J. \& Yagub, M. (Plenum, New York, 1965), p. 566

\bibitem{Berezinskii}
Berezinskii, V. L.
Destruction of long-range order in one-dimensional and two-dimensional systems possessing a continuous symmetry group. II. Quantum systems.
{\it Sov. Phys. JETP} {\bf 34,} 610 (1971).

\bibitem{KT}
Kosterlitz, J. M. \& Thouless, D. J.
Long range order and metastability in two dimensional solids and superfluids. (Application of dislocation theory).
{\it J. Phys. C} {\bf 5,} L124 (1972).
	
\bibitem{BMO}
Beasley, M. R., Mooij, J. E. \& Orlando, T. P.
Possibility of Vortex-Antivortex Pair Dissociation in Two-Dimensional Superconductors.
{\it Phys. Rev. Lett.} {\bf 42,} 1165 (1979).

\bibitem{Doniach}
Doniach, S. \& Huberman, B. A.
Topological Excitations in Two-Dimensional Superconductors.
{\it Phys. Rev. Lett.} {\bf 42,} 1169 (1979).

\bibitem{Minnhagen}
Minnhagen, P.
The two-dimensional Coulomb gas, vortex unbinding, and superfluid-superconducting films.
{\it Rev. Mod. Phys.} {\bf 59,} 1001 (1987).

\bibitem{Kosterlitz-Nelson}
Nelson, D. R. \& Kosterlitz, J. M.
Universal Jump in the Superfluid Density of Two-Dimensional Superfluids.
{\it Phys. Rev. Lett.} {\bf 39,} 1201 (1977).



\bibitem{Reyren}
Reyren, N. et al.
Superconducting Interfaces Between Insulating Oxides.
{\it Science} {\bf 317,} 1196 (2007).

\bibitem{CavigliaGate}
Caviglia, A. D., Gariglio, S., Reyren, N., Jaccard, D., Schneider, T., Gabay, M., Thiel, S., Hammerl, G., Mannhart, J. \& Triscone, J.-M.
Electric field control of the LaAlO$_3$/SrTiO$_3$ interface ground state
{\it Nature} {\bf 456,} 624 (2008).

\bibitem{Benfattobroad}
Benfatto, L., Castellani, C. \& Giamarchi, T.
Broadening of the Berezinskii-Kosterlitz-Thouless superconducting transition by inhomogeneity and finite-size effects.
{\it Phys. Rev. B} {\bf 80,} 214506 (2009).

\bibitem{BertSQUID}
Bert, J. A. et al.
Gate-tuned superfluid density at the superconducting LaAlO$_3$/SrTiO$_3$ interface.
{\it Phys. Rev. B} {\bf 86,} 060503(R) (2012).

\bibitem{ChangjiangLiu1}
Liu, C. et al.
Two-dimensional superconductivity and anisotropic transport at KTaO$_3$ (111) interfaces.
{\it  Science.} {\bf 371,} 716-721 (2021).

\bibitem{ChangjiangLiu2}
Liu, C. et al.
Tunable superconductivity and its origin at KTaO$_3$ interfaces.
{\it Nat. Commun.} {\bf 14,} 951 (2023).

\bibitem{YanwuXiePRL}
Chen, Z. et al.
Two-Dimensional Superconductivity at the LaAlO$_3$/KTaO$_3$(110) Heterointerface.
{\it Phys. Rev. Lett.} {\bf 126,} 026802 (2021).

\bibitem{HuaSOC}
Hua, X. et al.
Tunable two-dimensional superconductivity and spin-orbit coupling at the EuO/KTaO$_3$(110) interface.
{\it npj Quantum Mater.} {\bf 97,} 7 (2022).


\bibitem{KTOsuperfluid}
Mallik, S. et al.
Superfluid stiffness of a KTaO$_3$-based two-dimensional electron gas.
{\it Nat. Commun.} {\bf 13,} 4625 (2022).


\bibitem{Caprara}
Caprara, S. et al.
Multiband superconductivity and nanoscale inhomogeneity at oxide interfaces.
{\it Phys. Rev. B} {\bf 88,} 020504(R) (2013).

\bibitem{VendittiIV}
Venditti, G. et al.
Nonlinear I-V characteristics of two-dimensional superconductors: Berezinskii-Kosterlitz-Thouless physics versus inhomogeneity.
{\it Phys. Rev. B} {\bf 100,} 064506 (2019).

\bibitem{Scopigno}
Scopigno, N., Bucheli, D., Caprara, S., Biscaras, J., Bergeal, N., Lesueur, J. \& Grilli, M.
Phase Separation from Electron Confinement at Oxide Interfaces
{\it Phys. Rev. Lett.} {\bf 116,} 026804 (2016).

\bibitem{HuaStripe}
Hua, X. et al.
Superconducting stripes induced by ferromagnetic proximity in an oxide heterostructure.
{\it Nat. Phys.} {\bf 20,} 957 (2024).

\bibitem{Ahadi}
Arnault, E. G. et al.
Anisotropic superconductivity at KTaO$_3$(111) interfaces.
{\it Sci. Adv.} {\bf 9,} 1414 (2023).

\bibitem{CaZrO3Stoner}
Zhang, H. et al.
Magnetotransport evidence for the coexistence of two-dimensional superconductivity and ferromagnetism at (111)-oriented a-CaZrO$_3$/KTaO$_3$ interfaces.
{\it Nat. Commun.} {\bf 16,} 3035 (2025).

\bibitem{ZixiangLi}
Li, Z.-X., Kivelson, S. A. \& Lee, D.-H.
Theory of an infinitely anisotropic phase of a two-dimensional superconductor.
arXiv 2407.10269 (2024).


\bibitem{Zhang2DEG}
Zhang, H. et al.
High-Mobility Spin-Polarized Two-Dimensional Electron Gases at EuO/KTaO$_3$ Interfaces.
{\it Phys. Rev. Lett.} {\bf 121,} 116803 (2018).


\bibitem{MallikARPES}
Mallik, S. et al.
Electronic band structure of superconducting KTaO$_3$ (111) interfaces.
{\it APL Mater.} {\bf 11,} 121108 (2023).

\bibitem{HNformula}
Halperin, B. I. \& Nelson, D. R.
Resistive Transition in Superconducting Films.
{\it J. Low Temp. Phys.} {\bf 36,} 599 (1979).



\bibitem{BenfattoLa214}
Baity, P. G., Shi, X., Shi, Z., Benfatto, L. \& Dragana Popovi\'{c},
Effective two-dimensional thickness for the Berezinskii-Kosterlitz-Thouless-like transition in a highly underdoped La$_{2-x}$Sr$_x$CuO$_4$.
{\it Phys. Rev. B} {\bf 93,}  024519 (2016).

\bibitem{BrunoARPES}
Bruno, F. Y. et al.
Band structure and spin-orbital texture of the (111)-KTaO$_3$ 2D electron gas.
{\it Adv. Electron. Mater.} {\bf 5,} 1800860 (2019).


\bibitem{WeiLiRotation}
Zhang, G. et al.
Spontaneous rotational symmetry breaking in KTaO$_3$ heterointerface superconductors.
{\it Nat. Commun.} {\bf 14,} 3046 (2023).

\bibitem{HeBLMR}
He, P. et al.
Observation of Out-of-Plane Spin Texture in a SrTiO$_3$(111) Two-Dimensional Electron Gas.
{\it Phys. Rev. Lett.} {\bf 120,} 266802 (2018).

\bibitem{ChoeLAO-STO}
Choe, D. et al.
Gate-tunable giant nonreciprocal charge transport in noncentrosymmetric oxide interfaces.
{\it Nat. Commun.} {\bf 10,} 4510 (2019).

\bibitem{ZhangLightInduced}
Zhang, X. et al.
Light-induced giant enhancement of nonreciprocal transport at KTaO$_3$-based interfaces.
{\it  Nat. Commun.} {\bf 15,} 2992 (2024).

\bibitem{ItahashiSTO}
Itahashi, Y. M. et al.
Nonreciprocal transport in gate-induced polar superconductor SrTiO$_3$.
{\it Sci. Adv.} {\bf 9,} 9120 (2020).


\bibitem{WakatsukiMoS2}
Wakatsuki, R., Saito, Y., Hoshino, S., Itahashi, Y. M., Ideue, T., Ezawa, M., Iwasa, Y. \& Nagaosa, N.
Nonreciprocal charge transport in noncentrosymmetric superconductors.
{\it  Sci. Adv.} {\bf 3,} 1602390 (2017).


\bibitem{Hoshino}
Hoshino, S., Wakatsuki, R., Hamamoto, K. \& Nagaosa, N.
Nonreciprocal charge transport in two-dimensional noncentrosymmetric superconductors.
{\it Phys. Rev. B} {\bf 98,} 054510 (2018).


\bibitem{Masuko}
Masuko, M. et al.
Nonreciprocal charge transport in topological superconductor candidate Bi$_2$Te$_3$/PdTe$_2$ heterostructure.
{\it npj Quantum Mater.} {\bf 7,} 104 (2022).


\bibitem{Bi2Te3-FeTe}
Yasuda, K. et al.
Nonreciprocal charge transport at topological insulator/superconductor interface.
{\it Nat. Commun.} {\bf 10,} 2734 (2019).

\bibitem{BenfattoCoreEnegy}
Benfatto, L., Castellani, C. \& Giamarchi, T.
Kosterlitz-Thouless Behavior in Layered Superconductors: The Role of the Vortex Core Energy.
{\it Phys. Rev. Lett.} {\bf 98,} 117008 (2007).

\bibitem{YongNbN}
Yong, J., Lemberger, T. R., Benfatto, L., Ilin, K. \& Siegel, M.
Robustness of the Berezinskii-Kosterlitz-Thouless transition in ultrathin NbN films near the superconductor-insulator transition.
{\it Phys. Rev. B} {\bf 87,} 184505 (2013).

\bibitem{LAO-KTOARPES}
Chen, X. et al.
Orientation-dependent electronic structure in interfacial superconductors LaAlO$_3$/KTaO$_3$
{\it Nat. Commun.} {\bf 15,} 7704 (2024).

\bibitem{DomainWall}
Pai, Y.-Y. et al.
One-Dimensional Nature of Superconductivity at the LaAlO$_3$/SrTiO$_3$ Interface.
{\it Phys. Rev. Lett.} {\bf 120,} 147001 (2018).

\bibitem{WeiLiFM}
Ning, Z. et al.
Coexistence of Ferromagnetism and Superconductivity at KTaO$_3$ Heterointerfaces.
{\it Nano Lett.} {\bf 24,} 7134-7141 (2024).


\bibitem{PALee}
Michaeli, K., Potter, A. C. \& Lee, P. A.
Superconducting and Ferromagnetic Phases in SrTiO$_3$/LaAlO$_3$ Oxide Interface Structures: Possibility of Finite Momentum Pairing.
{\it Phys. Rev. Lett.} {\bf 108,} 117003 (2012).

\bibitem{Kozii}
Kozii, V. \& Fu, L.
Odd-Parity Superconductivity in the Vicinity of Inversion Symmetry Breaking in Spin-Orbit-Coupled Systems.
{\it Phys. Rev. Lett.} {\bf 115,} 207002 (2015).


\bibitem{WuNematicity}
Cheng, X. B. et al.
Electronic Nematicity in Interface Superconducting LAO/KTO(111).
{\it Phys. Rev. X.} {\bf 15,} 021018 (2025).

\bibitem{LiNSNO}
Xu, M. et al.
Anisotropic phase stiffness in infinite-layer nickelates superconductors.
{\it Nat. Commun.} {\bf 16,} 6780 (2025).

\bibitem{YanwuXiePRL2}
Sun, Y. et al.
Critical Thickness in Superconducting LaAlO$_3$/KTaO$_3$(111) Heterostructures.
{\it Phys. Rev. Lett.} {\bf 127,} 086804 (2021).

\end{thebibliography}

\begin{thebibliography}{99}
	
	
	
	


\bibitem{ChangjiangLiu1SI}
Liu, C. et al.
{\it Science} {\bf 371,} 716-721 (2021).


\bibitem{Zhang2DEGSI}
Zhang, H. et al.
{\it Phys. Rev. Lett.} {\bf 121,} 116803 (2018).


\bibitem{HuaSOCSI}
Hua, X. et al.
{\it npj Quantum Mater.} {\bf 97,} 7 (2022).

\bibitem{YanwuXieScience}
Chen, Z. et al.
{\it  Science} {\bf 372,} 721-724 (2021).

\bibitem{YanwuXiePRLSI}
Chen, Z. et al.
{\it Phys. Rev. Lett.} {\bf 126,} 026802 (2021).

\bibitem{HuaStripeSI}
Hua, X. et al.
{\it Nat. Phys.} {\bf 20,} 957 (2024).


\bibitem{CaZrO3StonerSI}
Zhang, H. et al.
{\it Nat. Commun.} {\bf 16,} 3035 (2025).

\bibitem{WeiLiFMSI}
Ning, Z. et al.
{\it Nano Lett.} {\bf 24,} 7134-7141 (2024).

\bibitem{Tinkham}
Tinkham, M. {\it Introduction to Superconductivity} (Dover
Publications, 2004)

\bibitem{2DHc}
Kozuka, Y., Kim, M., Bell, C., Kim, B. G., Hikita, Y. \& Hwang, H. Y.
{\it Nature} {\bf 462,} 487 (2009).


\bibitem{UPt3sixfold}
Krotkov, P. L. \& Mineev, V. P.
{\it Phys. Rev. B} {\bf 65,} 224506 (2002).

\bibitem{ALoriginal}
Aslamasov, L. G. \& Larkin, A. I.
{\it Phys. Lett. A.} {\bf 26,} 238 (1968).

\bibitem{ALCaprara}
Caprara, S., Grilli, M., Leridon, B. \& Lesueur, J.
{\it Phys. Rev. B} {\bf 72,} 104509 (2005).

\bibitem{BenfattoBroadening}
Benfatto, L., Castellani, C. \& Giamarchi, T.
{\it Phys. Rev. B} {\bf 80,} 214506 (2009).

\bibitem{BenfattoLa214SI}
Baity, P. G., Shi, X., Shi, Z., Benfatto, L. \& Popovi\'{c}, D.
{\it Phys. Rev. B} {\bf 93,}  024519 (2016).

\bibitem{HNformulaSI}
Halperin, B. I. \& Nelson, D. R.
{\it J. Low Temp. Phys.} {\bf 36,} 599 (1979).


\bibitem{Benfattolayered}
Benfatto, L., Castellani, C. \& Giamarchi, T.
{\it Phys. Rev. Lett.} {\bf 98,} 117008 (2007).

\bibitem{NbN}
Mondal, M., Kumar, S., Chand, M., Kamlapure, A., Saraswat, G., Seibold, G., Benfatto, L. \& Raychaudhuri, P.
{\it Phys. Rev. Lett.} {\bf 107,} 217003 (2011).

\bibitem{LiuNatComm}
Liu, C. et al.
{\it Nat. Commun.} {\bf 14,} 951 (2023).

\bibitem{BMOSI}
Beasley, M. R., Mooij, J. E. \& Orlando, T. P.
{\it Phys. Rev. Lett.} {\bf 42,} 1165 (1979).

\bibitem{YongNbNSI}
Yong, J., Lemberger, T. R., Benfatto, L., Ilin, K. \& Siegel, M.
{\it Phys. Rev. B} {\bf 87,} 184505 (2013).

\bibitem{CapraraSI}
Caprara, S. et al.
{\it Phys. Rev. B} {\bf 88,} 020504(R) (2013).

\bibitem{VendittiIVSI}
Venditti, G. et al.
{\it Phys. Rev. B} {\bf 100,} 064506 (2019).

\bibitem{KTOsuperfluidSI}
Mallik, S. et al.
{\it Nat. Commun.} {\bf 13,} 4625 (2022).

\bibitem{STOsuperfluid}
Singh, G. et al.
{\it Nat. Mater.} {\bf 18,} 948-954 (2019).

\bibitem{Bimodal}
Manca, N., Bothner, D., Monteiro, A. M. R. V. L., Davidovikj, D., Sa\u{g}lam, Y. G., Jenkins, M., Gabay, M., Steele, G. A. \& Caviglia, A. D.
{\it Phys. Rev. Lett.} {\bf 122,} 036801 (2019).

\bibitem{slowmotion}
Ganguly, R., Chaudhuri, D. Raychaudhuri, P. \& Benfatto, L.
{\it Phys. Rev. B} {\bf 91,}  054514 (2015).

\bibitem{percolation}
Caprara, S., Grilli, M., Benfatto, L. \& Castellani, C.
{\it Phys. Rev. B} {\bf 84,} 014514 (2011).

\bibitem{ReyrenSI}
Reyren, N. et al.
{\it Science} {\bf 317,} 1196 (2007).

\bibitem{fluxflow}
Ojha, S. K., Mandal, P., Kumar, S., Maity, J. \& Middey, S.
{\it Commun. Phys.} {\bf 6,} 126 (2023).

\bibitem{hotspot}
Gurevich, A. V. \& Mints, R. G.
{\it Rev. Mod. Phys.} {\bf 59,} 941 (1987).

\bibitem{JosephsonIV1}
Prawiroatmodjo, G. E. D. K., Trier, F., Christensen, D. V., Chen, Y., Pryds, N. \& Jespersen, T. S.
{\it Phys. Rev. B} {\bf 93,} 184504 (2016).

\bibitem{JosephsonIV2}
Hurand, S., Jouan, A., Lesne, E., Singh, G., Feuillet-Palma, C., Bibes, M., Barth\'{e}l\'{e}my, A., Lesueur, J. \& Bergeal, N.
{\it Phys. Rev. B} {\bf 99,} 104515 (2019).



\bibitem{LOinstability}
Larkin, A. \& Ovchinnikov, Y.
{\it Sov. Phys. JETP} {\bf 41,} 960 (1975).

\bibitem{Mo3Si}
Samoilov, A. V., Konczykowski, M., Yeh, N.-C., Berry, S. \& Tsuei, C. C.
{\it Phys. Rev. Lett.} {\bf 75,} 4118 (1995).


\bibitem{BiTeBr}
Ideue, T., Hamamoto, K., Koshikawa, S., Ezawa, M., Shimizu, S., Kaneko, Y., Tokura, Y., Nagaosa, N. \& Iwasa, Y.
{\it Nat. Phys.} {\bf 13,} 578-583 (2017).

\bibitem{Nonreciprocal}
Tokura, Y. \& Nagaosa, N.
{\it Nat. Commun.} {\bf 9,} 3740 (2018).

\bibitem{ChoeLAO-STOSI}
Choe, D. et al.
{\it Nat. Commun.} {\bf 10,} 4510 (2019).

\bibitem{WakatsukiMoS2SI}
Wakatsuki, R., Saito, Y., Hoshino, S., Itahashi, Y. M., Ideue, T., Ezawa, M., Iwasa, Y. \& Nagaosa, N.
{\it  Sci. Adv.} {\bf 3,} 1602390 (2017).


\bibitem{HoshinoSI}
Hoshino, S., Wakatsuki, R., Hamamoto, K. \& Nagaosa, N.
{\it Phys. Rev. B} {\bf 80,} 214506 (2018).


\bibitem{ItahashiSTOSI}
Itahashi, Y. M., Ideue, T., Saito, Y., Shimizu, S., Ouchi, T., Nojima, T. \& Iwasa, Y.
{\it Sci. Adv.} {\bf 9,} 9120 (2020).


\bibitem{MasukoSI}
Masuko, M. et al.
{\it npj Quantum Mater.} {\bf 7,} 104 (2022).

\bibitem{Bi2Te3-FeTeSI}
Yasuda, K., Yasuda, H., Liang, T., Yoshimi, R., Tsukazaki, A., Takahashi, K. S., Nagaosa, N., Kawasaki, M. \& Tokura, Y.
{\it Nat. Commun.} {\bf 10,} 2734 (2019).

\bibitem{FaxianXiu}
Zhang, E. et al.
{\it Nat. Commun.} {\bf 11,} 5634 (2020).

\bibitem{VortexReview}
Blatter, G. Feigel'man, M. Y. Geshkenbein, Y. B. Larkin, A. I. \& Vinokur, V. M.
{\it Rev. Mod. Phys.} {\bf 66,} 1125 (1994).

\bibitem{GalitskiLarkin}
Galitski, V. M. \& Larkin, A. I.
{\it Phys. Rev. B} {\bf 63,} 174506 (2001).

\bibitem{YBCO-MR1}
Rullier-Albenque, F., Alloul, H., Proust, C., Lejay, P., Forget, A. \& Colson, D.
{\it Phys. Rev. Lett.} {\bf 99,} 027003 (2007).

\bibitem{YBCO-MR2}
Rullier-Albenque, F., Alloul, H. \& Rikken, G.
{\it Phys. Rev. B} {\bf 84,} 014522 (2011).

\bibitem{Little}
Little, W. A.
{\it Phys. Rev.} {\bf 156,} 396 (1967).


\bibitem{Rogachev}
Rogachev, A., Bollinger, A. T. \& Bezryadin, A.
{\it Phys. Rev. Lett.} {\bf 94,} 017004 (2005).


\bibitem{CNLau}
Lau, C. N., Markovic, N., Bockrath, M., Bezryadin, A. \& Tinkham, M.
{\it Phys. Rev. Lett.} {\bf 87,} 217003 (2001).


\bibitem{LangerAmbegaokar}
Langer, J. S. \& Ambegaokar, V.
{\it Phys. Rev.} {\bf 164,} 498 (1967).

\bibitem{McCumberHalperin}
McCumber, D. E. \& Halperin, B. I.
{\it Phys. Rev. B} {\bf 1,} 1054-1070 (1970).


\end{thebibliography}
\end{document}